\newif\ifws
\ifws

\documentclass{article}

\usepackage{graphicx}        
\usepackage[dvipsnames]{xcolor}
\usepackage[normalem]{ulem}

\usepackage{hyperref}
\hypersetup{
    colorlinks,
    linkcolor={blue!80!black},
    citecolor={red!75!black},
    urlcolor={blue!80!black}
}

\title{Exploring Quantum Contextuality with the Quantum M\"obius--Escher--Penrose hypergraph}
\author{Mirko Navara \\
        Faculty of Electrical Engineering
Czech Technical University in Prague,
Technick\'a 2,
CZ-166~27 Prague 6,
Czech Republic
        }
\author{Karl Svozil \\
        Institute for Theoretical Physics,
TU Wien,  \\
Wiedner Hauptstrasse 8-10/136,
1040 Vienna,  Austria
        }

\date{\today}

\newcommand{
\new}[1]{{\color{magenta}#1}}
\newcommand{
\deleted}[1]{{\color{magenta}\sout{#1}}}

\begin{document}

\maketitle

\begin{abstract}
This paper presents the quantum M\"obius--Escher--Penrose hypergraph, drawing inspiration from paradoxical constructs such as the M\"obius strip and Penrose's `impossible objects'. The hypergraph is constructed using faithful orthogonal representations in Hilbert space, thereby embedding the graph within a quantum framework. Additionally, a quasi-classical realization is achieved through two-valued states and partition logic, leading to an embedding within a~Boolean algebra. This dual representation delineates the distinctions between classical and quantum embeddings, with a particular focus on contextuality, highlighted by violations of exclusivity and completeness, quantified through classical and quantum probabilities. The study also examines violations of Boole's conditions of possible experience using correlation polytopes, underscoring the inherent contextuality of the hypergraph. These results offer deeper insights into quantum contextuality and its intricate relationship with classical logic structures.
\end{abstract}

\else
\PassOptionsToPackage{dvipsnames}{xcolor}
\documentclass[
  preprint,
 showpacs,
 showkeys,
 preprintnumbers,
 amsmath,amssymb,
 aps,
  pra,
 longbibliography,
 floatfix,
 ]{revtex4-2}

\usepackage{mathptmx}

\usepackage{amssymb,amsthm,amsmath}

\usepackage{mathbbol}

\usepackage{tikz}
\usetikzlibrary{calc,math,decorations.shapes}
\usepackage {pgfplots}
\pgfplotsset {compat=1.8}
\usepackage{graphicx}

\usepackage{xcolor}
\usepackage[normalem]{ulem}

\usepackage{hyperref}
\hypersetup{
    colorlinks,
    linkcolor={blue},
    citecolor={red!75!black},
    urlcolor={blue}
}

\newcommand{
\new}[1]{{\color{magenta}#1}}
\newcommand{
\deleted}[1]{{\color{magenta}\sout{#1}}}

\begin{document}

\title{Exploring Quantum Contextuality with the Quantum M\"obius--Escher--Penrose hypergraph}

\author{Mirko Navara}
\email{navara@fel.cvut.cz}
\homepage{https://cmp.felk.cvut.cz/~navara}

\affiliation{Faculty of Electrical Engineering,
Czech Technical University in Prague,
Technick\'a 2,
CZ-166~27 Prague 6,
Czech Republic}

\author{Karl Svozil}
\email{karl.svozil@tuwien.ac.at}
\homepage{http://tph.tuwien.ac.at/~svozil}

\affiliation{Institute for Theoretical Physics,
TU Wien,
Wiedner Hauptstrasse 8-10/136,
1040 Vienna,  Austria}

\date{\today}

\begin{abstract}
This paper presents the quantum M\"obius--Escher--Penrose hypergraph, drawing inspiration from paradoxical constructs such as the M\"obius strip and Penrose's `impossible objects'. The hypergraph is constructed using faithful orthogonal representations in Hilbert space, thereby embedding the graph within a quantum framework. Additionally, a quasi-classical realization is achieved through two-valued states and partition logic, leading to an embedding within a~Boolean algebra. This dual representation delineates the distinctions between classical and quantum embeddings, with a particular focus on contextuality, highlighted by violations of exclusivity and completeness, quantified through classical and quantum probabilities. The paper also examines violations of Boole's conditions of possible experience using correlation polytopes, underscoring the inherent contextuality of the hypergraph. These results offer deeper insights into quantum contextuality and its intricate relationship with classical logic structures.
\end{abstract}


\maketitle

\newpage
\fi

\section{Introduction}

In this paper, we discuss a quantum analogue of the concept termed `impossible object' by Lionel Sharples Penrose and Roger Penrose~\cite{PENROSE_1958},
which was inspired by the paradoxical drawings of Maurits Cornelis Escher,
such as his  lithographs `Trappenhuis Stairs' and `Relativity'~\cite[plate 27]{Escher1954}.
These drawings depict structures that cannot exist in three-dimensional space,
leading to intriguing mathematical and artistic explorations.

The M\"obius strip, a related concept, has historically been represented in art and culture,
dating back to antiquity~\cite{cartwright-2016}.
With its single surface and edge, it challenges our conventional understanding of geometry
and has been a source of fascination and inspiration.

Building on these ideas, we introduce the M\"obius--Escher--Penrose hypergraph,
a structure depicted in Figure~\ref{2023-e-f1}.
This hypergraph was first introduced in a previous publication~\cite[Fig.~3, Eq.~(5)]{2023-navara-svozil},
where it was constructed using orthogonality hypergraphs.

The (hyper)edges of these hypergraphs represent the largest possible sets of mutually exclusive events,
which we will also refer to as contexts.
In quantum mechanics, all hyperedges (contexts) contain the same number of elements,
which can be identified with orthonormal bases, thereby forming uniform hypergraphs.
Contexts can intertwine when they contain more than two elements,
corresponding to orthonormal bases that share common elements.

The M\"obius--Escher--Penrose hypergraph
captures a very specific
structural connectivity of such intertwining contexts:
If one follows the context path (henceforth written in terms of the indices of their respective elements)
\(\{1,2,3\}-\{3,4,5\}-\{5,6,7\}-\{7,8,9\}-\{9,10,11\}-\{11,12,13\}-\{13,14,15\}-\{15,16,17\}-\{17,18,1\}-\{1,2,3\}\),
one realizes that these contexts `spiral back' to the original context \(\{1,2,3\}\) after a period of nine.
They are further intertwined by two additional contexts, \(\{2,8,14\}\) and \(\{4,10,16\}\).

If the two-valued measures separate elements on its edges, quasiclassical models of the propositional structure exist~\cite[Theorem~0]{kochen1}.
If these measures are derived from a quantum mechanical framework through orthogonal representations in Hilbert space,
the elements can be identified with vectors and their respective orthogonal projection operators.

In what follows we shall present  a more detailed exploration of the M\"obius--Escher--Penrose hypergraph,
highlighting its unique properties and its significance in understanding quantum contextuality
and the intricate relationship between classical logic structures and quantum mechanical systems.
In the next section, we elucidate how the hypergraph can be represented using quantum mechanics concepts such as vectors and operators.
The hypergraph is embedded into a quantum mechanical framework through orthogonal representations in Hilbert space.
Each edge of the hypergraph corresponds to an orthonormal basis,
ensuring a faithful embedding that respects the principles of quantum mechanics.

We introduce a classical counterpart to the quantum realization of the hypergraph.
By computing two-valued states and creating a partition logic,
we establish an embedding into a Boolean algebra.
This quasiclassical approach allows for a comparative analysis between classical and quantum embeddings,
highlighting the unique features and distinctions of quantum mechanical systems.

Next we demonstrate that certain classical aspects of the M\"obius--Escher--Penrose hypergraph involving quasicontexts
introduced in~\cite{2023-navara-svozil} violate the principle of exclusivity,
a cornerstone of classical probability theory.
These violations, however, are shown to be compatible with quantum mechanics.

Quantum contextuality can be expressed in terms of violations of George Boole's `conditions of possible experience'.
By translating binary value assignments into quantum mechanical expectations,
we show that the M\"obius--Escher--Penrose hypergraph violates these conditions in ways that classical physics cannot account for.
This exploration highlights the inherent contextuality of the hypergraph and the limitations
of classical explanations in capturing the vector space based probabilities of quantum systems.

\section{Quantum realization}

A faithful orthogonal representation (FOR) of a (hyper)graph involves labeling the graph with vectors, where adjacency corresponds to orthogonality \cite{lovasz-79,GroetschelLovaszSchrijver1986,Portillo-2015}. The associated quantum observables consist of orthogonal (self-adjoint) projection operators formed by dyadic products of these vectors. Thus, hypergraph edges correspond to contexts identified with orthonormal bases of Hilbert space.

Without loss of generality, a faithful orthogonal representation of the M\"obius--Escher--Penrose hypergraph depicted in Figure~\ref{2023-e-f1} can be constructed by beginning with an orthogonal tripod of vectors
$
v_4 = \begin{pmatrix} \sqrt{2}, 0, 1 \end{pmatrix}^\intercal
$,
$
v_{16} = \begin{pmatrix} -1, \sqrt{3}, \sqrt{2} \end{pmatrix}^\intercal
$, and
$
v_{10} = \begin{pmatrix} -1, -\sqrt{3}, \sqrt{2} \end{pmatrix}^\intercal
$,
where $^\intercal$ stands for transposition.
This tripod is rotated around the $z$-axis
$
\vert z\rangle = \begin{pmatrix} 0,0,1 \end{pmatrix}^\intercal
$ by the angle
\begin{equation}
\label{2024-e-alpha}
\alpha
=
2 \cot ^{-1}\left(\sqrt{\frac{11}{9}+\frac{1}{81}
   \sqrt[3]{2262816-69984 \sqrt{69}}+\frac{2^{5/3}}{9}
   \sqrt[3]{97+3 \sqrt{69}}}\right),
\end{equation}
resulting in the tripod of vectors
$v_2$, $v_{14}$, and $v_8$, respectively.
The construction progresses by taking the successive cross products
$v_3  =  v_4 \times v_2          $,
$v_{15} =  v_{16} \times v_{14}        $,
$v_9 =  v_{10} \times v_8          $,
$v_5  =   v_3 \times v_4         $,
$v_{17} =   v_{15} \times v_{16}       $,
$v_{11} =   v_9 \times v_{10}        $,
$v_1 =   v_3 \times v_2          $,
$v_{13} =   v_{15} \times v_{14}       $,
$v_7 =   v_9 \times v_8          $,
$v_6 =   v_7 \times v_5          $,
$v_{12} =   v_{13} \times v_{11}       $,  and
$v_{18} =   v_1 \times v_{17}        $.

\begin{figure}
\begin{center}
\begin{tabular}{ccc}
\resizebox{.46\textwidth}{!}{
\begin{tikzpicture} [scale=2]
\tikzstyle{every path}=[line width=2pt]

    \coordinate (9) at (90:2);    
    \coordinate (8) at (30:2);    
    \coordinate (2) at (-30:2);   
    \coordinate (3) at (-90:2);   
    \coordinate (4) at (-150:2);  
    \coordinate (10) at (150:2);   

    \node at ($(9) + (90:0.3)$) {9};
    \node at ($(8) + (30:0.3)$) {8};
    \node at ($(2) + (-30:0.3)$) {2};
    \node at ($(3) + (-90:0.3)$) {3};
    \node at ($(4) + (-150:0.3)$) {4};
    \node at ($(10) + (150:0.3)$) {10};

    \coordinate (7) at ($(9)!0.5!(8)$); 
    \coordinate (14) at ($(8)!0.5!(2)$); 
    \coordinate (1) at ($(2)!0.5!(3)$); 
    \coordinate (5) at ($(3)!0.5!(4)$); 
    \coordinate (16) at ($(4)!0.5!(10)$); 
    \coordinate (11) at ($(10)!0.5!(9)$); 


    \node at ($(7) + (60:0.3)$) {7};
    \node at ($(14) + (0:0.3)$) {14};
    \node at ($(1) + (-60:0.3)$) {1};
    \node at ($(5) + (-120:0.3)$) {5};
    \node at ($(16) + (180:0.3)$) {16};
    \node at ($(11) + (120:0.3)$) {11};

    \coordinate (15) at (90:1);    
    \node at ($(15) + (90:0.3)$) {15};

    \draw[orange]  (16) --  (15);
    \coordinate (17) at ($(16)!0.5!(15)$); 
    \node at ($(17) + (170:0.3)$) {17};

    \draw[green]  (15) -- (14);
    \coordinate (13) at ($(15)!0.5!(14)$); 
    \node at ($(13) + (10:0.3)$) {13};

    \draw[gray]  (7) -- (5);
    \coordinate (6) at ($(7)!0.5!(5)$); 
    \node at ($(6) + (0:0.3)$) {6};

    \draw[brown]  (17) -- (1);
    \coordinate (18) at ($(17)!0.7!(1)$); 
    \node at ($(18) + (190:0.3)$) {18};

    \draw[violet] (11) .. controls (100:0.5) .. (13);
    \path (11) .. controls (100:0.5) .. (13) coordinate[pos=0.14] (12);

    \node at ($(12) + (75:0.3)$) {12};

    \draw [red] (1) -- (3);
    \draw [red] (1) -- (2);

    \draw [blue] (3) -- (5);
    \draw [blue] (4) -- (5);


    \draw [magenta] (7) -- (9);
    \draw [magenta] (7) -- (8);

    \draw [cyan] (9) -- (11);
    \draw [cyan] (10) -- (11);

    \draw [olive] (4) -- (10);
    \draw [teal] (8) -- (2);

    \draw [ultra thick,gray,dotted] (5) -- (11);
    \draw [ultra thick,gray,dotted] (1) -- (7);

    \fill[red    ] (1) circle  (3pt);
    \fill[red    ] (2) circle  (3pt);
    \fill[red    ] (3) circle  (3pt);
    \fill[blue   ] (4) circle  (3pt);
    \fill[blue   ] (5) circle  (3pt);
    \fill[gray   ] (6) circle  (3pt);
    \fill[magenta] (7) circle  (3pt);
    \fill[magenta] (8) circle  (3pt);
    \fill[cyan   ] (9) circle  (3pt);
    \fill[cyan   ] (10) circle (3pt);
    \fill[cyan   ] (11) circle (3pt);
    \fill[violet ] (12) circle (3pt);
    \fill[green  ] (13) circle (3pt);
    \fill[green  ] (14) circle (3pt);
    \fill[orange ] (15) circle (3pt);
    \fill[orange ] (16) circle (3pt);
    \fill[orange ] (17) circle (3pt);
    \fill[brown  ] (18) circle (3pt);

    \fill[brown  ] (1) circle  (2pt);
    \fill[teal   ] (2) circle  (2pt);
    \fill[blue   ] (3) circle  (2pt);
    \fill[olive  ] (4) circle  (2pt);
    \fill[gray   ] (5) circle  (2pt);

    \fill[gray   ] (7) circle  (2pt);
    \fill[teal   ] (8) circle  (2pt);
    \fill[magenta] (9) circle  (2pt);
    \fill[olive  ] (10) circle (2pt);
    \fill[violet ] (11) circle (2pt);

    \fill[violet ] (13) circle (2pt);
    \fill[teal   ] (14) circle (2pt);
    \fill[green  ] (15) circle (2pt);
    \fill[olive  ] (16) circle (2pt);
    \fill[brown  ] (17) circle (2pt);

\end{tikzpicture}
}
&$\qquad$&
\resizebox{.46\textwidth}{!}{\begin{tikzpicture}  [scale=1] 
\tikzstyle{every path}=[line width=2pt]

\tikzmath{\a = 1; \b = 2; \c = 0;}

\newdimen\ms
\ms=0.05cm

\tikzstyle{c3}=[circle,inner sep={\ms/8},minimum size=8*\ms]
\tikzstyle{c2}=[circle,inner sep={\ms/8},minimum size=5*\ms]
\tikzstyle{c1}=[circle,inner sep={\ms/8},minimum size=0.8*\ms]

\coordinate (1) at ({2*( \a +  \a + \b ) + 2*\c},{( \a +  \a + \b ) + 2*\c});
\coordinate (3) at ({( \a +  \a + \b )},0);
\coordinate (2) at ($(1)!0.5!(3)$);
\coordinate (5) at ({0-2*\c},{( \a +  \a + \b )+2*\c});
\coordinate (4) at ($(3)!0.5!(5)$);
\coordinate (13) at ({2*( \a +  \a + \b )-\a+\c},{( \a +  \a + \b )+\c});
\coordinate (15) at ({( \a +  \a + \b )},\a);
\coordinate (14) at ($(13)!0.5!(15)$);
\coordinate (17) at ({\a-\c},{( \a +  \a + \b )+\c});
\coordinate (16) at ($(15)!0.5!(17)$);
\coordinate (7) at ({2*( \a +  \a + \b )-2*\a},{( \a +  \a + \b )});
\coordinate (9) at ({( \a +  \a + \b )},{2*\a});
\coordinate (8) at ($(7)!0.5!(9)$);
\coordinate (11) at ({2*\a},{( \a +  \a + \b )});
\coordinate (10) at ($(9)!0.5!(11)$);

 \coordinate (12) at ($(11)!0.5!(13)$);
 \coordinate (18) at ($(17)!0.5!(1)$);

\draw [red] (1) -- (3);
\draw [blue] (3) -- (5);

\draw [green] (13) -- (15);
\draw [orange] (15) -- (17);

\draw [magenta] (7) -- (9);
\draw [cyan] (9) -- (11);

\draw [olive] (4) -- (10);
\draw [teal] (8) -- (2);

\draw [ultra thick,gray,dotted] (5) -- (11);
\draw [ultra thick,gray,dotted] (1) -- (7);

\draw [gray] (5) .. controls (\a,{1.6*( \a +  \a + \b )}) and ({ 2*( \a +  \a + \b ) - 2} ,{1.6*( \a +  \a + \b )}) .. (7)
                coordinate [pos=0.3] (6);

\draw [brown] (17) .. controls ({ ( \a +  \a + \b ) - 2},{1.5*( \a +  \a + \b )}) and ({  ( \a +  \a + \b ) + 2} ,{1.5*( \a +  \a + \b )}) .. (1)
                coordinate [pos=0.8] (18);

\draw [violet] (11) .. controls ({ ( \a +  \a + \b ) - 2},{1.3*( \a +  \a + \b )}) and ({ ( \a +  \a + \b ) + 2} ,{1.3*( \a +  \a + \b )}) .. (13)
                coordinate [pos=0.2] (12);

\draw (1)  coordinate[c3,fill=red     ,label={below right: $1$}];
\draw (2)  coordinate[c3,fill=red     ,label={below right: $2$}];
\draw (3)  coordinate[c3,fill=red     ,label={below: $3$}];
\draw (4)  coordinate[c3,fill=blue    ,label={below: $4$}];
\draw (5)  coordinate[c3,fill=blue    ,label={below: $5$}];
\draw (6)  coordinate[c3,fill=gray    ,label={above left: $6$}];
\draw (7)  coordinate[c3,fill=magenta ,label={below: $7$}];
\draw (8)  coordinate[c3,fill=magenta ,label={below: $8$}];
\draw (9)  coordinate[c3,fill=cyan    ,label={below: $9$}];
\draw (10) coordinate[c3,fill=cyan    ,label={below: $10$}];
\draw (11) coordinate[c3,fill=cyan    ,label={below: $11$}];
\draw (12) coordinate[c3,fill=violet  ,label={below: $12$}];
\draw (13) coordinate[c3,fill=green   ,label={below: $13$}];
\draw (14) coordinate[c3,fill=green   ,label={below: $14$}];
\draw (15) coordinate[c3,fill=orange  ,label={below: $15$}];
\draw (16) coordinate[c3,fill=orange  ,label={below: $16$}];
\draw (17) coordinate[c3,fill=orange  ,label={below: $17$}];
\draw (18) coordinate[c3,fill=brown   ,label={above: $18$}];

\draw (1)  coordinate[c2,fill=brown  ];
\draw (2)  coordinate[c2,fill=teal   ];
\draw (3)  coordinate[c2,fill=blue   ];
\draw (4)  coordinate[c2,fill=olive  ];
\draw (5)  coordinate[c2,fill=gray   ];

\draw (7)  coordinate[c2,fill=gray   ];
\draw (8)  coordinate[c2,fill=teal   ];
\draw (9)  coordinate[c2,fill=magenta];
\draw (10) coordinate[c2,fill=olive  ];
\draw (11) coordinate[c2,fill=violet ];

\draw (13) coordinate[c2,fill=violet ];
\draw (14) coordinate[c2,fill=teal   ];
\draw (15) coordinate[c2,fill=green  ];
\draw (16) coordinate[c2,fill=olive  ];
\draw (17) coordinate[c2,fill=brown  ];

\end{tikzpicture}}
\end{tabular}
\end{center}
\caption{\label{2023-e-f1}
Equivalent representations of the M\"obius--Escher--Penrose hypergraph include smooth or straight lines to denote contexts,
while dotted lines represent pseudocontexts~\cite{2023-navara-svozil}.
The (index) labels $i$ stand for vectors $v_i$ or, more generally, elements $a_i$.
}
\end{figure}

\section{Quasiclassical realization}

A~\emph{state} $m$, representing the probabilistic or truth-value assignments, must
yield  the sum of~$1$ over all elements of a~context $\{a_1,\ldots,a_n\}$:
\begin{equation}\label{eq.context}
m(a_1) + \cdots + m(a_n)
= 1.
\end{equation}

A~\emph{two-valued state} has the range of $\{0,1\}$. Therefore, it selects exactly one $i$ such that ${m(a_i)=1}$.
This requirement is split to two elementary properties:
\begin{itemize}
    \item[(i)] \emph{Exclusivity}: There is at most one $i$ such that $m(a_i)=1$,
    \item[(ii)] \emph{Completeness}: There exists $i$ such that $m(a_i)=1$.
\end{itemize}

A~\emph{partition logic} is based on a set of partitions of a given finite set (without loss of generality, set of natural numbers).
Each partition is interpreted as a Boolean algebra whose atoms correspond to elementary propositions of that algebra.
In a final step, all of these algebras (corresponding to the partitions) are~\emph{pasted} together to form the partition logic.
As an elementary example, consider the set of two partitions
$
\Bigl\{\bigl\{\{1,2\},\{3,4\}\bigr\},\bigl\{\{1,3\},\{2,4\}\bigr\}\Bigr\}
$
of the four-element set $\{1,2,3,4\}$.
These partitions correspond to two Boolean algebras, each isomorphic to $2^2$:
one with atoms $\{1,2\}$ and $\{3,4\}$, and the other with atoms $\{1,3\}$ and $\{2,4\}$.
Their pasting~\cite{nav:91} forms a \emph{horizontal sum}, often referred to as a `Chinese lantern', with two common elements:
the maximal element $\{1,2,3,4\}$ and the minimal element $\emptyset$.
The order relation is defined by set-theoretic inclusion.
Such a logic can still be considered `classical' and noncontextual because it allows a faithful embedding into a Boolean algebra.
However, it also exhibits \emph{complementarity}, as the measurement of one subalgebra entails no knowledge of the other.

A quasiclassical embedding of the M\"obius--Escher--Penrose hypergraph is achieved by computing all $12$ two-valued states,
noticing that they are separating~\cite[Theorem~0]{kochen1}; and by constructing a partition logic
based on the set of indices of nonvanishing states associated with each hypergraph element~\cite{svozil-2001-eua}.
Table~\ref{2024-e-t1} enumerates the vector labels in both quasiclassical terms, represented by partition elements,
and quantum terms, represented by vectors.

\begin{table}
\caption{\label{2024-e-t1}Partition logic and vector label representations
of the  M\"obius--Escher--Penrose hypergraph.
$R\bigl(   \vert z\rangle , \alpha \bigr)$
 stands for the rotation matrix around the $z$-axis $\begin{pmatrix} 0,0,1 \end{pmatrix}^\intercal$
by the angle
$\alpha$ defined in~(\ref{2024-e-alpha}).
}
\begin{ruledtabular}
\begin{tabular}{ccc}
$a_i$ & partition element & vector $v_i$\\
\hline
$a_{1 }$& $\{1,2,3\}$            & $v_3 \times v_2          $                                                     \\
$a_{2 }$& $\{4,5,6,7\}$          & $R\bigl(   \vert z\rangle , \alpha \bigr) v_4    $        \\
$a_{3 }$& $\{8,9,10,11,12\}$     & $v_4 \times v_2          $                                                     \\
$a_{4 }$& $\{1,4,5,6\}$          & $\begin{pmatrix} \sqrt{2}, 0, 1 \end{pmatrix}^\intercal$                               \\
$a_{5 }$& $\{2,3,7\}$            & $v_3 \times v_4         $                                                      \\
$a_{6 }$& $\{1,4,5,8,9,10\}$     & $v_7 \times v_5          $                                                     \\
$a_{7 }$& $\{6,11,12\}$          & $v_9 \times v_8          $                                                     \\
$a_{8 }$& $\{1,2,8,9\}$          & $R\bigl(   \vert z\rangle , \alpha \bigr) v_{10}    $     \\
$a_{9 }$& $\{3,4,5,7,10\}$       & $v_{10} \times v_8          $                                                  \\
$a_{10}$& $\{2,8,9,11\}$         & $\begin{pmatrix} -1, -\sqrt{3}, \sqrt{2} \end{pmatrix}^\intercal$                      \\
$a_{11}$& $\{1,6,12\}$           & $v_9 \times v_{10}        $                                                    \\
$a_{12}$& $\{2,3,4,8,10,11\}$    & $v_{13} \times v_{11}       $                                                  \\
$a_{13}$& $\{5,7,9\}$            & $v_{15} \times v_{14}       $                                                  \\
$a_{14}$& $\{3,10,11,12\}$       & $R\bigl(   \vert z\rangle , \alpha \bigr) v_{16}    $     \\
$a_{15}$& $\{1,2,4,6,8\}$        & $v_{16} \times v_{14}        $                                                 \\
$a_{16}$& $\{3,7,10,12\}$        & $\begin{pmatrix} -1, \sqrt{3}, \sqrt{2} \end{pmatrix}^\intercal$                       \\
$a_{17}$& $\{5,9,11\}$           & $v_{15} \times v_{16}       $                                                  \\
$a_{18}$& $\{4,6,7,8,10,12\}$    & $v_1 \times v_{17}        $                                                    \\
\end{tabular}
\end{ruledtabular}
\end{table}

The construction of the quasiclassical faithful (homomorphic) embedding into a Boolean algebra \( 2^{12} \)~\cite{ZirlSchl-65}
is facilitated by a separating~\cite[Theorem~0]{kochen1} set of two-valued states.
The Travis matrix~\cite{travis-mt-62,greechie-66-PhD}
is a~matrix whose columns correspond to vertices of a~hypergraph (vectors in FOR) and rows correspond to two-valued states.
The matrix contains entries $0$ and $1$ depending on the evaluation of the vertex by the respective state.
The Travis matrix for the $12$ two-valued states on the hypergraph in Figure~\ref{2023-e-f1}
is enumerated in Table~\ref{2024-e-t2}.

\begin{table}
\caption{\label{2024-e-t2}Twelve two-valued (binary) states on the M\"obius--Escher--Penrose hypergraph.
}
\begin{ruledtabular}
\begin{tabular}{c|ccccccccccccccccccccccccccc}
$a_i$/\#&$a_{1}$&$a_{2}$&$a_{3}$&$a_{4}$&$a_{5}$&$a_{6}$&$a_{7}$&$a_{8}$&$a_{9}$&$a_{10}$&$a_{11}$&$a_{12}$&$a_{13}$&$a_{14}$&$a_{15}$&$a_{16}$&$a_{17}$&$a_{18}$\\
\hline
1  & 1& 0& 0& 1& 0& 1& 0& 1& 0& 0& 1& 0& 0& 0& 1& 0& 0& 0  \\
2  & 1& 0& 0& 0& 1& 0& 0& 1& 0& 1& 0& 1& 0& 0& 1& 0& 0& 0  \\
3  & 1& 0& 0& 0& 1& 0& 0& 0& 1& 0& 0& 1& 0& 1& 0& 1& 0& 0  \\
4  & 0& 1& 0& 1& 0& 1& 0& 0& 1& 0& 0& 1& 0& 0& 1& 0& 0& 1  \\
5  & 0& 1& 0& 1& 0& 1& 0& 0& 1& 0& 0& 0& 1& 0& 0& 0& 1& 0  \\
6  & 0& 1& 0& 1& 0& 0& 1& 0& 0& 0& 1& 0& 0& 0& 1& 0& 0& 1  \\
7  & 0& 1& 0& 0& 1& 0& 0& 0& 1& 0& 0& 0& 1& 0& 0& 1& 0& 1  \\
8  & 0& 0& 1& 0& 0& 1& 0& 1& 0& 1& 0& 1& 0& 0& 1& 0& 0& 1  \\
9  & 0& 0& 1& 0& 0& 1& 0& 1& 0& 1& 0& 0& 1& 0& 0& 0& 1& 0  \\
10 & 0& 0& 1& 0& 0& 1& 0& 0& 1& 0& 0& 1& 0& 1& 0& 1& 0& 1  \\
11 & 0& 0& 1& 0& 0& 0& 1& 0& 0& 1& 0& 1& 0& 1& 0& 0& 1& 0  \\
12 & 0& 0& 1& 0& 0& 0& 1& 0& 0& 0& 1& 0& 0& 1& 0& 1& 0& 1  \\
\end{tabular}
\end{ruledtabular}
\end{table}

\section{Contextuality by violation of exclusivity}

Let us consider an orthogonal basis $\{a_1, \ldots, a_n\}$.
As noted earlier, it represents a~\emph{context}, a maximal
set of events that can be simultaneously tested in one experiment.
Their states---representing the probabilistic or truth-value assignments---satisfy
\begin{equation}\label{eq.context1}
m(a_1) + \cdots + m(a_n)
= 1.
\end{equation}
for any state~$m$. 

Equations such as~\eqref{eq.context} or ~\eqref{eq.context1} 
can hold also for collections of  elements
called pseudocontexts~\cite{2023-navara-svozil}
which are complementary, and, in the quantum context, do not satisfy any orthogonality relation.
They contain collections of elements in a
hypergraph that have a total probability sum equal to that of other collections of elements in the same hypergraph.
Because they are complementary, elements in such pseudocontexts are not necessarily mutually exclusive.
Their probability measures do not necessarily sum to one.

In what follows we shall specify this condition by exemplifying
two pseudocontexts as two sets of three vectors, where the probability sum of the first three equals the probability
sum of the second three for all quantum states.

%
%
%
Classically, for two-valued states, the two pseudocontexts $\{5,11,17\}$ and $\{1,7,13\}$ (henceforth written in terms of the indices of their elements) obey exclusivity:
if one of their elements has value $1$, the others must be $0$.
This property arises because the pairs of their elements form a true-implies-false (TIFS) gadget~\cite{2018-minimalYIYS},
specifically a Specker bug~\cite[Fig.~1, p.~182]{kochen2} (reprinted in Ref.~\cite{specker-ges}].
This can be verified by inspecting the set of two-valued states in Table~\ref{2024-e-t2}
or by proof by contradiction:
Assume both elements have value $1$ and follow admissibility until a contradiction is reached,
such as all elements of a context having value $0$ or two elements in a context having value $1$.

The pseudocontexts do not satisfy completeness, as all elements can have value $0$.
Consequently, we derive an upper bound on the classical probabilistic or truth-value assignments on pseudocontexts:
\begin{equation}
m(a_5) + m(a_{11}) + m(a_{17}) = m(a_1) + m(a_7) + m(a_{13}) \le 1.
\label{2014-e-bfapc}
\end{equation}

These bounds are maximally violated by the quantum probabilities for states perpendicular to the rotation axis
$\vert z\rangle = \begin{pmatrix} 0,0,1 \end{pmatrix}^\intercal$,
as the (multiple, identical) eigenvalue of $E_5+E_{11}+E_{17}=E_1+E_7+E_{13}$
with $E_i = \vert v_i \rangle \langle v_i \vert /  \langle v_i \vert   v_i \rangle$,
for the $\vert x\rangle = \begin{pmatrix} 1,0,0 \end{pmatrix}^\intercal$ and $\vert y\rangle = \begin{pmatrix} 0,1,0 \end{pmatrix}^\intercal$ axes,
is
\begin{equation}
\begin{aligned}
&\langle x \vert E_5+E_{11}+E_{17} \vert x \rangle = \langle x \vert E_1+E_7+E_{13}  \vert x \rangle
\\
&\quad = \langle y \vert E_5+E_{11}+E_{17} \vert y \rangle = \langle y \vert E_1+E_7+E_{13}  \vert y \rangle
\\
&\quad =\frac{1}{6} \left(10-10 \sqrt[3]{\frac{2}{3 \sqrt{69}-11}}+2^{2/3} \sqrt[3]{3 \sqrt{69}-11}\right)
\\
&\quad \approx 1.43016
.
\end{aligned}
\label{2014-e-bfapc-q}
\end{equation}
This two-dimensional form of quantum contextuality is
in-between Hardy-type paradoxes~\cite{Cabello-2013-Hardylike,2018-minimalYIYS}
that operate with a single TIFS resulting in violations for a one-dimensional subspace of Hilbert space,
and Yu and Oh's state-independent proof of {K}ochen-{S}pecker Theorem~\cite{Yu-2012,cabello2021contextuality}.

The TIFS pairs
$(2,10)$,
$(4,14)$
and $(8,16)$,
require the `paradoxical periodic closure' of the hypergraph.

\section{Contextuality by violation of {B}oole's `Conditions of Possible Experience'}

In addition to the primary constraints on probabilities \( p_i \) of events---namely,
that they should not be negative or greater than one---there are ``other conditions'' that will, as Boole pointed out~\cite[p.~229]{Boole-62},
\emph{``be capable of expression by
equations or inequations reducible to the general (linear) form
\(a_1p_1+a_2p_2+\cdots a_n p_n+a \ge 0, \)
$a_1, a_2, \ldots a_n, a$
being numerical constants which differ for the different conditions in
question.''}
This applies also to joint probabilities, and affine---in particular, linear---transformations thereof.

In what follows we shall transcribe binary value assignments into expectations and operator values
by the affine transformation
$ A = 1 - 2 s $ which, for each context or hyperedge, produces two values $1$ and a~single value~$-1$, corresponding
to $s$ being $0$ and~$1$, respectively.
Quantum mechanically this translates into Householder transformations~\cite{svozil-2021-hh}
$\mathbf{A} = \mathbb{1}_3 - 2 \vert x \rangle \langle x \vert $,
where $\vert x \rangle$ is the unit vector corresponding to the quantum state
$\vert x \rangle \langle x \vert$
associated with a two-valued state~$s$ \cite{Klyachko-2008,cabello:210401}.
Note that the way it is constructed $\mathbf{A}$ has a unit eigenvector $\vert x \rangle$ with eigenvalue~$-1$;
more explicitly, $\mathbf{A}\vert x \rangle =   \mathbb{1}_3 \vert x \rangle - 2 \vert x \rangle \underbrace{\langle x  \vert x \rangle}_{1}
= - \vert x \rangle$.
The (multiple) eigenvalue~$1$ is obtained for any orthonormal basis spanning the (hyper)plane orthogonal to $\vert x \rangle$:
Take, for instance, any unit vector $\vert y \rangle$ with $ \langle x \vert y \rangle =0$.
Then  $\mathbf{A}\vert y \rangle =   \mathbb{1}_3 \vert y \rangle - 2 \vert x \rangle \underbrace{\langle x  \vert y \rangle}_{0}
=   \vert y \rangle$.

A systematic route to
{B}oole's `conditions of possible experience'~\cite{Boole-62,Pit-94}
is via (correlation) polytopes: First the `extreme' cases are computed
which are then encoded into vectors.
In a~second step `mixed' classical probabilities or correlations are obtained
by a~convex sum over those extreme cases, and vectors.
The latter convex linear combination of vectors forms a correlation polytope.
In a third and final stage, the polytope is represented
by an equivalent representation in term of its faces---its hull~\cite{ziegler}.
The transcription of vertex to the latter representation
is called the hull problem.
The classical bounds---{B}oole's
`conditions of possible experience'---can then be
identified with the face (in)equalities
obtained from solving the hull problem~\cite{froissart-81,pitowsky-86},
also, in a generalized form, for  intertwining (pasted~\cite{nav:91}) contexts~\cite{svozil-2001-cesena,Klyachko-2008}.

Although the polytope method is systematic, the particular choice of the correlations yielding deviations with {B}oole's
conditions of possible experience appears to be heuristic and \textit{ad hoc.}
Let us, for instance, form 12 vectors $w_i$ with the index $1 \le i \le 12$ referring to the $i$th  two-valued state
enumerated in Table~\ref{2024-e-t2}.
The components of $w_i$  are the products of two  elements of the same context---the ones that intertwine with the next context, following
a M\"obius--Escher--Penrose path $1-3-5-7-9-11-13-15-17-1$ of contexts:
\begin{equation}
w_i =
\begin{pmatrix}
A_{1 } A_{3 }  ,
A_{3 } A_{5 }  ,
A_{5 } A_{7 }  ,
A_{7 } A_{9 }  ,
A_{9 } A_{11}  ,
A_{11} A_{13}  ,
A_{13} A_{15}  ,
A_{15} A_{17}  ,
A_{17} A_{1 }
\end{pmatrix}^\intercal_i
\label{2024-e-booleconfig}
\end{equation}
Components $w_1 , \ldots , w_{12}$ can be interpreted as nine-dimensional vertices of a convex polytope
$\lambda_1 w_1 + \cdots  +\lambda_{12} w_{12}$
with the convex sum $\lambda_1  + \cdots  + \lambda_{12} =1$ for
$0\le \lambda_i \le 1$, and $1 \le i \le 12$.

\subsection{Hull equalities}

Solving the hull problem yields 30 faces~\cite{cdd-pck,pycddlib},
among them two equalities
\begin{equation}
\begin{split}
 \langle A_{3}A_{5 } \rangle + \langle A_{9}A_{11 } \rangle + \langle A_{15}A_{17 } \rangle = -1 \,,  \\
 \langle A_{1}A_{3 } \rangle + \langle A_{7}A_{9 } \rangle  + \langle A_{13}A_{15 } \rangle = -1 \,.  \\
\end{split}
\label{2024-e-Booleequal}
\end{equation}
Classically, they are  consequences of exclusivity and completeness (admissibility):
Exactly one of the elements of the contexts
$\{4,10,16\}$
and
$\{2,8,14\}$
has to be assigned the value $s=1$ and thus $A=-1$, and two elements $s=0$ and thus $A=1$.
As a consequence the product of the other two elements of the three context intertwining with, say,
$\{4,10,16\}$, needs to be once $1 \times 1=1$ and twice $-1 \times 1= -1$.

This argument does not need the full structure of the M\"obius--Escher--Penrose hypergraph:
A~`pruned' subset consisting of four contexts and depicted in Figure~\ref{2023-e-fsubs4con} suffices.
The quantum double of this pruned configuration is
$
 \mathbf{B}_{4}\mathbf{B}_{5 }
+   \mathbf{B}_{6}\mathbf{B}_{7}
+   \mathbf{B}_{8}\mathbf{B}_{9}
= - \mathbb{1}_3
$
which is satisfied also for
$
\mathbf{A}_3\mathbf{A}_5
+\mathbf{A}_5\mathbf{A}_9
+\mathbf{A}_9\mathbf{A}_{11}
= - \mathbb{1}_3
$, as well as for
$\mathbf{A}_{11}\mathbf{A}_{15}
+\mathbf{A}_{15}\mathbf{A}_{17}
+\mathbf{A}_{17}\mathbf{A}_1
= - \mathbb{1}_3
$.
Hence, no discrepancy with classical expectations,
and, therefore, no quantum contextuality
can be derived from the two hull equalities~(\ref{2024-e-Booleequal}).

Nevertheless, these equalities yield an intuitive understanding
of the pseudocontexts~\cite{2023-navara-svozil}:
because two such gadgets as the one depicted in Figure~\ref{2023-e-fsubs4con}, `tied together'
at three elements---in this case $3$, $9$, and $15$---with equal sums, require
the respective sum of the `open ends'
$\{1,7,13\}$ and
$\{5,11,17\}$ are also equal.

\begin{figure}
\begin{center}
\resizebox{.35\textwidth}{!}{
\begin{tikzpicture}
\tikzstyle{every path}=[line width=2.5pt]

\draw [red] (0,0) -- (10,0);

\draw [blue] (0,0) -- (0,6);
\draw [orange] (5,0) -- (5,6);
\draw [green] (10,0) -- (10,6);

\fill [red] (0,0) circle (6pt) node[below=0.23cm] {\color{black}1};
\fill [red] (5,0) circle (6pt) node[below=0.23cm] {\color{black}2};
\fill [red] (10,0) circle (6pt) node[below=0.23cm] {\color{black}3};

\fill [blue] (0,0) circle (4pt);
\fill [orange] (5,0) circle (4pt);
\fill [green] (10,0) circle (4pt);

\fill [blue] (0,3) circle (4pt) node[right=0.15cm] {\color{black}4};
\fill [orange] (5,3) circle (4pt) node[right=0.15cm] {\color{black}6};
\fill [green] (10,3) circle (4pt) node[right=0.15cm] {\color{black}8};

\fill [blue] (0,6) circle (4pt) node[right=0.15cm] {\color{black}5};
\fill [orange] (5,6) circle (4pt) node[right=0.15cm] {\color{black}7};
\fill [green] (10,6) circle (4pt) node[right=0.15cm] {\color{black}9};

\end{tikzpicture}
}
\end{center}
\caption{\label{2023-e-fsubs4con}
Gadget formed from a subset of the M\"obius--Escher--Penrose hypergraph reproducing equalities~\eqref{2024-e-Booleequal}.
The operator-valued equality from the hull computation in this pruned configuration is
$\langle B_{4}B_{5 } \rangle + \langle B_{6}B_{7} \rangle + \langle B_{8}B_{9} \rangle = -1$,
with $B_{i} = 1-2 v_i$.
}
\end{figure}

\subsection{Hull inequalities}

The remaining 28 hull inequalities can be grouped to collections of descending number of summands:
\begin{equation}
\begin{aligned}
1:-1&\le 2 \langle A_1A_3\rangle - 2 \langle A_3A_5\rangle - \langle A_5A_7\rangle + 2 \langle A_7A_9\rangle - 2 \langle A_9A_{11}\rangle + \langle A_{11}A_{13}\rangle - \langle A_{17}A_{1} \rangle ,\\
2:\phantom{-}1&\ge  2 \langle A_1A_3\rangle - 2 \langle A_3A_5\rangle + \langle A_5A_7\rangle + 2 \langle A_7A_9\rangle - 2 \langle A_9A_{11}\rangle + \langle A_{11}A_{13}\rangle - \langle A_{17}A_{1}\rangle ,\\
3:-1&\le 2 \langle A_1A_3\rangle - 2 \langle A_3A_5\rangle + \langle A_5A_7\rangle - 2 \langle A_9A_{11}\rangle + \langle A_{11}A_{13}\rangle - \langle A_{17}A_{1}\rangle ,\\
4:-1&\le  - 2 \langle A_1A_3\rangle + \langle A_5A_7\rangle - 2 \langle A_7A_9\rangle + 2 \langle A_9A_{11}\rangle - \langle A_{11}A_{13}\rangle + \langle A_{17}A_{1}\rangle ,\\
5:-3&\le 2 \langle A_3A_5\rangle - \langle A_5A_7\rangle + 2 \langle A_9A_{11}\rangle + \langle A_{11}A_{13}\rangle + \langle A_{17}A_{1}\rangle ,\\
6:-3&\le 2 \langle A_3A_5\rangle - \langle A_5A_7\rangle + 2 \langle A_7A_9\rangle + \langle A_{11}A_{13}\rangle + \langle A_{17}A_{1} \rangle ,\\
7:-3&\le 2 \langle A_1A_3\rangle - \langle A_5A_7\rangle + 2 \langle A_7A_9\rangle + \langle A_{11}A_{13}\rangle + \langle A_{17}A_{1} \rangle ,\\
8:-1&\le 2 \langle A_1A_3\rangle - 2 \langle A_3A_5\rangle + \langle A_5A_7\rangle - \langle A_{11}A_{13}\rangle - \langle A_{17}A_{1}\rangle ,\\
9:-1&\le  - 2 \langle A_1A_3\rangle + 2 \langle A_3A_5\rangle - \langle A_5A_7\rangle - \langle A_{11}A_{13}\rangle + \langle A_{17}A_{1} \rangle ,\\
10:-1&\le \langle A_5A_7\rangle - 2 \langle A_7A_9\rangle + 2 \langle A_9A_{11}\rangle - \langle A_{11}A_{13}\rangle - \langle A_{17}A_{1}\rangle ,\\
11:-1&\le  - \langle A_5A_7\rangle + 2 \langle A_7A_9\rangle - 2 \langle A_9A_{11}\rangle + \langle A_{11}A_{13}\rangle - \langle A_{17}A_{1}\rangle ,\\
12-15:-1&\le \{ - 2 \langle A_3A_5\rangle + \langle A_5A_7\rangle - \langle A_{11}A_{13}\rangle + \langle A_{17}A_{1} \rangle , - 2 \langle A_1A_3\rangle + \langle A_5A_7\rangle \\
 & \qquad - \langle A_{11}A_{13}\rangle + \langle A_{17}A_{1} \rangle, \langle A_5A_7\rangle - 2 \langle A_9A_{11}\rangle + \langle A_{11}A_{13}\rangle - \langle A_{17}A_{1}\rangle , \\
 & \qquad \langle A_5A_7\rangle - 2 \langle A_7A_9\rangle + \langle A_{11}A_{13}\rangle - \langle A_{17}A_{1} \rangle \},\\
16-19:-1&\le \{ \langle A_5A_7\rangle + \langle A_{11}A_{13}\rangle + \langle A_{17}A_{1} \rangle, \langle A_5A_7\rangle - \langle A_{11}A_{13}\rangle - \langle A_{17}A_{1} \rangle ,\\
  & \qquad  - \langle A_5A_7\rangle - \langle A_{11}A_{13}\rangle + \langle A_{17}A_{1} \rangle ,  - \langle A_5A_7\rangle + \langle A_{11}A_{13}\rangle - \langle A_{17}A_{1} \rangle \}, \\
20,21:-1&\le \{\langle A_3A_5\rangle - \langle A_7A_9\rangle + \langle A_9A_{11} \rangle , \langle A_1A_3\rangle - \langle A_3A_5\rangle + \langle A_7A_9 \rangle \} ,\\
22-24:\phantom{-}0&\le  \{ - \langle A_3A_5\rangle - \langle A_9A_{11}\rangle , - \langle A_1A_3\rangle - \langle A_9A_{11}\rangle ,  - \langle A_1A_3\rangle - \langle A_7A_9 \rangle \} ,\\
25-28:-1&\le \{ \langle A_3A_5\rangle ,  \langle A_1A_3\rangle ,  \langle A_7A_9 \rangle ,   \langle A_9A_{11}\rangle \} .
\end{aligned}
\label{2024-e-bbti}
\end{equation}
Heuristically, the `longest' such summations are the `most likely' to be violated quantum mechanically,
whereas the `shortest' are, according to Boole, just rules
from the requirement for probability  `of being positive proper fractions'~\cite[p.~229]{Boole-62}.
Table~\ref{2024-e-contextualities} enumerates the  violations of Boole's generalized (for multiple intertwining context) conditions of possible experience:
\begin{table}
\caption{\label{2024-e-contextualities}Contextuality by violation of Boole's generalized
(for multiple intertwining context) conditions of possible experience.}
\begin{ruledtabular}
\begin{tabular}{cccccccccccccccccccccccccccc}
\# & eigenvalue of maximal violation & eigenvector \\
from Eq.~(\ref{2024-e-bbti})& (numerical) & (numerical) \\
\hline
1  & -3 & $\begin{pmatrix}0.981557, 0.173328, -0.0806446\end{pmatrix}^\intercal$  \\
5  & -3.89807 & $\begin{pmatrix}0.491309, -0.837518, -0.239123\end{pmatrix}^\intercal$  \\
6  & -3.89807 & $\begin{pmatrix}0.95734, -0.162235, -0.239123\end{pmatrix}^\intercal$  \\
7  & -2.26894 & $\begin{pmatrix}0.195441, -0.683898, -0.702913\end{pmatrix}^\intercal$  \\
12  & -1.89807 & $\begin{pmatrix}0.970966, 0.00672715, 0.239123\end{pmatrix}^\intercal$  \\
13  & -1.89807 & $\begin{pmatrix}0.619169, 0.747964, 0.239123\end{pmatrix}^\intercal$  \\
14  & -1.89807 & $\begin{pmatrix}0.479657, 0.844245, -0.239123\end{pmatrix}^\intercal$  \\
15  & -2.12744 & $\begin{pmatrix}0.553247, -0.811751, 0.187022\end{pmatrix}^\intercal$  \\
16  & -1.36373 & $\begin{pmatrix}0.0343782, 0.426292, -0.903932\end{pmatrix}^\intercal$  \\
17  & -1.36373 & $\begin{pmatrix}0.351991, -0.242918, -0.903932\end{pmatrix}^\intercal$  \\
18  & -1.36373 & $\begin{pmatrix}0.0343782, 0.426292, -0.903932\end{pmatrix}^\intercal$  \\
19  & -1.36373 & $\begin{pmatrix}0.386369, 0.183374, 0.903932\end{pmatrix}^\intercal$  \\
20  & -1.64944 & $\begin{pmatrix}0.428768, -0.903415, 0\end{pmatrix}^\intercal$  \\
21  & -1.64944 & $\begin{pmatrix}0.996764, -0.0803837, 0\end{pmatrix}^\intercal$  \\
23  & -0.64944 & $\begin{pmatrix}0.567996, 0.823031, 0\end{pmatrix}^\intercal$
\end{tabular}
\end{ruledtabular}
\end{table}

\section{Chromatic contextuality}
If contexts are considered representations of maximal empirical knowledge about a quantum state and are identified with maximal observables \cite[Satz~8, p.~221f]{v-neumann-31} (for a contemporary review, see Halmos \cite[\S84, p.171f]{halmos-vs}), then the study of chromatic numbers and hypergraph colorings \cite{svozil-2021-chroma,svozil-2025-color} becomes highly significant. In particular, each color corresponds to a distinct measurement outcome when measuring a context associated with a maximal observable. This observable's nondegenerate spectrum includes all orthogonal projection operators corresponding to the vectors labeling the hypergraph vertices.

A brute-force computation yields \(3! \times 3 = 18\) possible colorings of the M\"obius--Escher--Penrose hypergraph,
as enumerated in Table~\ref{2024-e-t-chroma} and depicted in Figure~\ref{2023-e-f-color}.
The factor \(3!\) accounts for permutations of colors, which do not provide any structural information.
Consequently, the remaining three distinct colorings (up to permutations) represent the fundamental chromatic modes of the hypergraph.

\begin{table}
\caption{\label{2024-e-t-chroma}Three non-isomorphic (with respect to permutations of colors) colorings (up to permutations of the colors) of the M\"obius--Escher--Penrose hypergraph.
}
\begin{ruledtabular}
\begin{tabular}{c|ccccccccccccccccccccccccccc}
$a_i$/\#&$a_{1}$&$a_{2}$&$a_{3}$&$a_{4}$&$a_{5}$&$a_{6}$&$a_{7}$&$a_{8}$&$a_{9}$&$a_{10}$&$a_{11}$&$a_{12}$&$a_{13}$&$a_{14}$&$a_{15}$&$a_{16}$&$a_{17}$&$a_{18}$\\
\hline
1 &  1& 2& 3& 1& 2& 1& 3& 1& 2& 3& 1& 3& 2& 3& 1& 2& 3& 2 \\
2 &  1& 2& 3& 2& 1& 2& 3& 1& 2& 1& 3& 1& 2& 3& 1& 3& 2& 3 \\
3 &  1& 2& 3& 2& 1& 3& 2& 3& 1& 3& 2& 1& 3& 1& 2& 1& 3& 2
\end{tabular}
\end{ruledtabular}
\end{table}

\begin{figure}
\begin{center}
\begin{tabular}{ccccc}
\resizebox{.27\textwidth}{!}{
\begin{tikzpicture} [scale=2]
\tikzstyle{every path}=[line width=2pt]

    \coordinate (9) at (90:2);    
    \coordinate (8) at (30:2);    
    \coordinate (2) at (-30:2);   
    \coordinate (3) at (-90:2);   
    \coordinate (4) at (-150:2);  
    \coordinate (10) at (150:2);   

    \node at ($(9) + (90:0.3)$) {\color{black}9};
    \node at ($(8) + (30:0.3)$) {\color{black}8};
    \node at ($(2) + (-30:0.3)$) {\color{black}2};
    \node at ($(3) + (-90:0.3)$) {\color{black}3};
    \node at ($(4) + (-150:0.3)$) {\color{black}4};
    \node at ($(10) + (150:0.3)$) {\color{black}10};

    \coordinate (7) at ($(9)!0.5!(8)$); 
    \coordinate (14) at ($(8)!0.5!(2)$); 
    \coordinate (1) at ($(2)!0.5!(3)$); 
    \coordinate (5) at ($(3)!0.5!(4)$); 
    \coordinate (16) at ($(4)!0.5!(10)$); 
    \coordinate (11) at ($(10)!0.5!(9)$); 


    \node at ($(7) + (60:0.3)$) {\color{black}7};
    \node at ($(14) + (0:0.3)$) {\color{black}14};
    \node at ($(1) + (-60:0.3)$) {\color{black}1};
    \node at ($(5) + (-120:0.3)$) {\color{black}5};
    \node at ($(16) + (180:0.3)$) {\color{black}16};
    \node at ($(11) + (120:0.3)$) {\color{black}11};

    \coordinate (15) at (90:1);    
    \node at ($(15) + (90:0.3)$) {\color{black}15};

    \draw[orange]  (16) --  (15);
    \coordinate (17) at ($(16)!0.5!(15)$); 
    \node at ($(17) + (170:0.3)$) {\color{black}17};

    \draw[green]  (15) -- (14);
    \coordinate (13) at ($(15)!0.5!(14)$); 
    \node at ($(13) + (10:0.3)$) {\color{black}13};

    \draw[gray]  (7) -- (5);
    \coordinate (6) at ($(7)!0.5!(5)$); 
    \node at ($(6) + (0:0.3)$) {\color{black}6};

    \draw[brown]  (17) -- (1);
    \coordinate (18) at ($(17)!0.7!(1)$); 
    \node at ($(18) + (190:0.3)$) {\color{black}18};

    \draw[violet] (11) .. controls (100:0.5) .. (13);
    \path (11) .. controls (100:0.5) .. (13) coordinate[pos=0.14] (12);

    \node at ($(12) + (75:0.3)$) {\color{black}12};

    \draw [red] (1) -- (3);
    \draw [red] (1) -- (2);

    \draw [blue] (3) -- (5);
    \draw [blue] (4) -- (5);


    \draw [magenta] (7) -- (9);
    \draw [magenta] (7) -- (8);

    \draw [cyan] (9) -- (11);
    \draw [cyan] (10) -- (11);

    \draw [olive] (4) -- (10);
    \draw [teal] (8) -- (2);

    \draw [ultra thick,gray,dotted] (5) -- (11);
    \draw [ultra thick,gray,dotted] (1) -- (7);

    \fill[red, decorate, decoration=zigzag ] (1) circle  (4pt);
    \fill[green] (2) circle  (4pt);
    \node[blue, fill, minimum width=14pt, minimum height=14pt] at  (3) {};
    \fill[red, decorate, decoration=zigzag  ] (4) circle  (4pt);
    \fill[green] (5) circle  (4pt);
    \fill[red, decorate, decoration=zigzag  ] (6) circle  (4pt);
    \node[blue, fill, minimum width=14pt, minimum height=14pt] at  (7) {};
    \fill[red, decorate, decoration=zigzag  ] (8) circle  (4pt);
    \fill[green] (9) circle  (4pt);
    \node[blue, fill, minimum width=14pt, minimum height=14pt] at  (10) {};
    \fill[red, decorate, decoration=zigzag  ] (11) circle (4pt);
    \node[blue, fill, minimum width=14pt, minimum height=14pt] at  (12) {};
    \fill[green] (13) circle (4pt);
    \node[blue, fill, minimum width=14pt, minimum height=14pt] at  (14) {};
    \fill[red, decorate, decoration=zigzag  ] (15) circle (4pt);
    \fill[green] (16) circle (4pt);
    \node[blue, fill, minimum width=14pt, minimum height=14pt] at  (17) {};
    \fill[green] (18) circle (4pt);

\end{tikzpicture}
}
&$\qquad$&
\resizebox{.27\textwidth}{!}{
\begin{tikzpicture} [scale=2]
\tikzstyle{every path}=[line width=2pt]

    \coordinate (9) at (90:2);    
    \coordinate (8) at (30:2);    
    \coordinate (2) at (-30:2);   
    \coordinate (3) at (-90:2);   
    \coordinate (4) at (-150:2);  
    \coordinate (10) at (150:2);   

    \node at ($(9) + (90:0.3)$) {\color{black}9};
    \node at ($(8) + (30:0.3)$) {\color{black}8};
    \node at ($(2) + (-30:0.3)$) {\color{black}2};
    \node at ($(3) + (-90:0.3)$) {\color{black}3};
    \node at ($(4) + (-150:0.3)$) {\color{black}4};
    \node at ($(10) + (150:0.3)$) {\color{black}10};

    \coordinate (7) at ($(9)!0.5!(8)$); 
    \coordinate (14) at ($(8)!0.5!(2)$); 
    \coordinate (1) at ($(2)!0.5!(3)$); 
    \coordinate (5) at ($(3)!0.5!(4)$); 
    \coordinate (16) at ($(4)!0.5!(10)$); 
    \coordinate (11) at ($(10)!0.5!(9)$); 


    \node at ($(7) + (60:0.3)$) {\color{black}7};
    \node at ($(14) + (0:0.3)$) {\color{black}14};
    \node at ($(1) + (-60:0.3)$) {\color{black}1};
    \node at ($(5) + (-120:0.3)$) {\color{black}5};
    \node at ($(16) + (180:0.3)$) {\color{black}16};
    \node at ($(11) + (120:0.3)$) {\color{black}11};

    \coordinate (15) at (90:1);    
    \node at ($(15) + (90:0.3)$) {\color{black}15};

    \draw[orange]  (16) --  (15);
    \coordinate (17) at ($(16)!0.5!(15)$); 
    \node at ($(17) + (170:0.3)$) {\color{black}17};

    \draw[green]  (15) -- (14);
    \coordinate (13) at ($(15)!0.5!(14)$); 
    \node at ($(13) + (10:0.3)$) {\color{black}13};

    \draw[gray]  (7) -- (5);
    \coordinate (6) at ($(7)!0.5!(5)$); 
    \node at ($(6) + (0:0.3)$) {\color{black}6};

    \draw[brown]  (17) -- (1);
    \coordinate (18) at ($(17)!0.7!(1)$); 
    \node at ($(18) + (190:0.3)$) {\color{black}18};

    \draw[violet] (11) .. controls (100:0.5) .. (13);
    \path (11) .. controls (100:0.5) .. (13) coordinate[pos=0.14] (12);

    \node at ($(12) + (75:0.3)$) {\color{black}12};

    \draw [red] (1) -- (3);
    \draw [red] (1) -- (2);

    \draw [blue] (3) -- (5);
    \draw [blue] (4) -- (5);


    \draw [magenta] (7) -- (9);
    \draw [magenta] (7) -- (8);

    \draw [cyan] (9) -- (11);
    \draw [cyan] (10) -- (11);

    \draw [olive] (4) -- (10);
    \draw [teal] (8) -- (2);

    \draw [ultra thick,gray,dotted] (5) -- (11);
    \draw [ultra thick,gray,dotted] (1) -- (7);

    \fill[red, decorate, decoration=zigzag  ] (1) circle  (4pt);
    \fill[green] (2) circle  (4pt);
    \node[blue, fill, minimum width=14pt, minimum height=14pt] at  (3) {};
    \fill[green] (4) circle  (4pt);
    \fill[red, decorate, decoration=zigzag  ] (5) circle  (4pt);
    \fill[green] (6) circle  (4pt);
    \node[blue, fill, minimum width=14pt, minimum height=14pt] at  (7) {};
    \fill[red, decorate, decoration=zigzag  ] (8) circle  (4pt);
    \fill[green] (9) circle  (4pt);
    \fill[red, decorate, decoration=zigzag  ] (10) circle (4pt);
    \node[blue, fill, minimum width=14pt, minimum height=14pt] at  (11) {};
    \fill[red, decorate, decoration=zigzag  ] (12) circle (4pt);
    \fill[green] (13) circle (4pt);
    \node[blue, fill, minimum width=14pt, minimum height=14pt] at  (14) {};
    \fill[red, decorate, decoration=zigzag  ] (15) circle (4pt);
    \node[blue, fill, minimum width=14pt, minimum height=14pt] at  (16) {};
    \fill[green] (17) circle (4pt);
    \node[blue, fill, minimum width=14pt, minimum height=14pt] at  (18) {};

\end{tikzpicture}
}
&$\qquad$&
\resizebox{.27\textwidth}{!}{
\begin{tikzpicture} [scale=2]
\tikzstyle{every path}=[line width=2pt]

    \coordinate (9) at (90:2);    
    \coordinate (8) at (30:2);    
    \coordinate (2) at (-30:2);   
    \coordinate (3) at (-90:2);   
    \coordinate (4) at (-150:2);  
    \coordinate (10) at (150:2);   

    \node at ($(9) + (90:0.3)$) {\color{black}9};
    \node at ($(8) + (30:0.3)$) {\color{black}8};
    \node at ($(2) + (-30:0.3)$) {\color{black}2};
    \node at ($(3) + (-90:0.3)$) {\color{black}3};
    \node at ($(4) + (-150:0.3)$) {\color{black}4};
    \node at ($(10) + (150:0.3)$) {\color{black}10};

    \coordinate (7) at ($(9)!0.5!(8)$); 
    \coordinate (14) at ($(8)!0.5!(2)$); 
    \coordinate (1) at ($(2)!0.5!(3)$); 
    \coordinate (5) at ($(3)!0.5!(4)$); 
    \coordinate (16) at ($(4)!0.5!(10)$); 
    \coordinate (11) at ($(10)!0.5!(9)$); 


    \node at ($(7) + (60:0.3)$) {\color{black}7};
    \node at ($(14) + (0:0.3)$) {\color{black}14};
    \node at ($(1) + (-60:0.3)$) {\color{black}1};
    \node at ($(5) + (-120:0.3)$) {\color{black}5};
    \node at ($(16) + (180:0.3)$) {\color{black}16};
    \node at ($(11) + (120:0.3)$) {\color{black}11};

    \coordinate (15) at (90:1);    
    \node at ($(15) + (90:0.3)$) {\color{black}15};

    \draw[orange]  (16) --  (15);
    \coordinate (17) at ($(16)!0.5!(15)$); 
    \node at ($(17) + (170:0.3)$) {\color{black}17};

    \draw[green]  (15) -- (14);
    \coordinate (13) at ($(15)!0.5!(14)$); 
    \node at ($(13) + (10:0.3)$) {\color{black}13};

    \draw[gray]  (7) -- (5);
    \coordinate (6) at ($(7)!0.5!(5)$); 
    \node at ($(6) + (0:0.3)$) {\color{black}6};

    \draw[brown]  (17) -- (1);
    \coordinate (18) at ($(17)!0.7!(1)$); 
    \node at ($(18) + (190:0.3)$) {\color{black}18};

    \draw[violet] (11) .. controls (100:0.5) .. (13);
    \path (11) .. controls (100:0.5) .. (13) coordinate[pos=0.14] (12);

    \node at ($(12) + (75:0.3)$) {\color{black}12};

    \draw [red] (1) -- (3);
    \draw [red] (1) -- (2);

    \draw [blue] (3) -- (5);
    \draw [blue] (4) -- (5);


    \draw [magenta] (7) -- (9);
    \draw [magenta] (7) -- (8);

    \draw [cyan] (9) -- (11);
    \draw [cyan] (10) -- (11);

    \draw [olive] (4) -- (10);
    \draw [teal] (8) -- (2);

    \draw [ultra thick,gray,dotted] (5) -- (11);
    \draw [ultra thick,gray,dotted] (1) -- (7);

    \fill[red, decorate, decoration=zigzag  ] (1) circle  (4pt);
    \fill[green] (2) circle  (4pt);
    \node[blue, fill, minimum width=14pt, minimum height=14pt] at  (3) {};
    \fill[green] (4) circle  (4pt);
    \fill[red, decorate, decoration=zigzag  ] (5) circle  (4pt);
    \node[blue, fill, minimum width=14pt, minimum height=14pt] at  (6) {};
    \fill[green] (7) circle  (4pt);
    \node[blue, fill, minimum width=14pt, minimum height=14pt] at  (8) {};
    \fill[red, decorate, decoration=zigzag  ] (9) circle  (4pt);
    \node[blue, fill, minimum width=14pt, minimum height=14pt] at  (10) {};
    \fill[green] (11) circle (4pt);
    \fill[red, decorate, decoration=zigzag  ] (12) circle (4pt);
    \node[blue, fill, minimum width=14pt, minimum height=14pt] at  (13) {};
    \fill[red, decorate, decoration=zigzag  ] (14) circle (4pt);
    \fill[green] (15) circle (4pt);
    \fill[red, decorate, decoration=zigzag  ] (16) circle (4pt);
    \node[blue, fill, minimum width=14pt, minimum height=14pt] at  (17) {};
    \fill[green] (18) circle (4pt);

\end{tikzpicture}
}
\\
(1)&&(2)&&(3)
\end{tabular}
\end{center}
\caption{\label{2023-e-f-color}
Three non-isomorphic (with respect to permutations of colors) colorings of the M\"obius--Escher--Penrose hypergraph, as enumerated in Table~\ref{2024-e-t-chroma}. The colors are represented by distinct shapes: circles, stars, and squares, respectively.
}
\end{figure}

Any such coloring is also a valid coloring of the pseudocontexts
\(\{1,7,13\}\) and \(\{5,11,17\}\), as all three colors appear in each pseudocontext for all legal colorings of the hypergraph.
This is particularly remarkable because any `reduced' two-valued state---based on colorings, where
a single color is mapped to the value \(1\) and the remaining two colors to \(0\)---effectively
transforms the pseudocontexts into classical contexts.
That is, for such reduced measures on pseudocontexts, the sum of the measures must equal one.

Three of the two-valued measures---specifically, state numbers 4, 8, and 10, as listed in Table~\ref{2024-e-t2} and illustrated
in Figure~\ref{2023-e-f-color-tws}---cannot be extended to a 3-coloring.
Their sums vanish on the pseudocontexts.
Consequently, removing them in the set-theoretic embedding, where a partition logic is represented in Table~\ref{2024-e-t1},
results in two complete partitions---and thus, contexts.

Therefore, with this reduced set of two-valued states we obtain classical predictions that yield
\begin{equation}
m(a_1)+m(a_7)+m(a_{13})=m(a_5)+m(a_{11})+m(a_{17})=1,
\label{2014-e-spcredtvs}
\end{equation}
for the sum of states of each set---as compared to the upper bound of one in Equation~(\ref{2014-e-bfapc}).

\begin{figure}
\begin{center}
\begin{tabular}{ccccc}
\resizebox{.27\textwidth}{!}{
\begin{tikzpicture} [scale=2]
\tikzstyle{every path}=[line width=2pt]

    \coordinate (9) at (90:2);    
    \coordinate (8) at (30:2);    
    \coordinate (2) at (-30:2);   
    \coordinate (3) at (-90:2);   
    \coordinate (4) at (-150:2);  
    \coordinate (10) at (150:2);   

    \node at ($(9) + (90:0.3)$) {\color{black}9};
    \node at ($(8) + (30:0.3)$) {\color{black}8};
    \node at ($(2) + (-30:0.3)$) {\color{black}2};
    \node at ($(3) + (-90:0.3)$) {\color{black}3};
    \node at ($(4) + (-150:0.3)$) {\color{black}4};
    \node at ($(10) + (150:0.3)$) {\color{black}10};

    \coordinate (7) at ($(9)!0.5!(8)$); 
    \coordinate (14) at ($(8)!0.5!(2)$); 
    \coordinate (1) at ($(2)!0.5!(3)$); 
    \coordinate (5) at ($(3)!0.5!(4)$); 
    \coordinate (16) at ($(4)!0.5!(10)$); 
    \coordinate (11) at ($(10)!0.5!(9)$); 


    \node at ($(7) + (60:0.3)$) {\color{black}7};
    \node at ($(14) + (0:0.3)$) {\color{black}14};
    \node at ($(1) + (-60:0.3)$) {\color{black}1};
    \node at ($(5) + (-120:0.3)$) {\color{black}5};
    \node at ($(16) + (180:0.3)$) {\color{black}16};
    \node at ($(11) + (120:0.3)$) {\color{black}11};

    \coordinate (15) at (90:1);    
    \node at ($(15) + (90:0.3)$) {\color{black}15};

    \draw[orange]  (16) --  (15);
    \coordinate (17) at ($(16)!0.5!(15)$); 
    \node at ($(17) + (170:0.3)$) {\color{black}17};

    \draw[green]  (15) -- (14);
    \coordinate (13) at ($(15)!0.5!(14)$); 
    \node at ($(13) + (10:0.3)$) {\color{black}13};

    \draw[gray]  (7) -- (5);
    \coordinate (6) at ($(7)!0.5!(5)$); 
    \node at ($(6) + (0:0.3)$) {\color{black}6};

    \draw[brown]  (17) -- (1);
    \coordinate (18) at ($(17)!0.7!(1)$); 
    \node at ($(18) + (190:0.3)$) {\color{black}18};

    \draw[violet] (11) .. controls (100:0.5) .. (13);
    \path (11) .. controls (100:0.5) .. (13) coordinate[pos=0.14] (12);

    \node at ($(12) + (75:0.3)$) {\color{black}12};

    \draw [red] (1) -- (3);
    \draw [red] (1) -- (2);

    \draw [blue] (3) -- (5);
    \draw [blue] (4) -- (5);


    \draw [magenta] (7) -- (9);
    \draw [magenta] (7) -- (8);

    \draw [cyan] (9) -- (11);
    \draw [cyan] (10) -- (11);

    \draw [olive] (4) -- (10);
    \draw [teal] (8) -- (2);

    \draw [ultra thick,gray,dotted] (5) -- (11);
    \draw [ultra thick,gray,dotted] (1) -- (7);

    \fill[green ] (1) circle  (4pt);
    \fill[red, decorate, decoration=zigzag] (2) circle  (4pt);
    \fill[green] (3) circle  (4pt);
    \fill[red, decorate, decoration=zigzag] (4) circle  (4pt);
    \fill[green] (5) circle  (4pt);
    \fill[red, decorate, decoration=zigzag] (6) circle  (4pt);
    \fill[green] (7) circle  (4pt);
    \fill[green ] (8) circle  (4pt);
    \fill[red, decorate, decoration=zigzag] (9) circle  (4pt);
    \fill[green] (10) circle (4pt);
    \fill[green ] (11) circle (4pt);
    \fill[red, decorate, decoration=zigzag] (12) circle (4pt);
    \fill[green] (13) circle (4pt);
    \fill[green] (14) circle (4pt);
    \fill[red, decorate, decoration=zigzag ] (15) circle (4pt);
    \fill[green] (16) circle (4pt);
    \fill[green] (17) circle (4pt);
    \fill[red, decorate, decoration=zigzag] (18) circle (4pt);

\end{tikzpicture}
}
&$\qquad$&
\resizebox{.27\textwidth}{!}{
\begin{tikzpicture} [scale=2]
\tikzstyle{every path}=[line width=2pt]

    \coordinate (9) at (90:2);    
    \coordinate (8) at (30:2);    
    \coordinate (2) at (-30:2);   
    \coordinate (3) at (-90:2);   
    \coordinate (4) at (-150:2);  
    \coordinate (10) at (150:2);   

    \node at ($(9) + (90:0.3)$) {\color{black}9};
    \node at ($(8) + (30:0.3)$) {\color{black}8};
    \node at ($(2) + (-30:0.3)$) {\color{black}2};
    \node at ($(3) + (-90:0.3)$) {\color{black}3};
    \node at ($(4) + (-150:0.3)$) {\color{black}4};
    \node at ($(10) + (150:0.3)$) {\color{black}10};

    \coordinate (7) at ($(9)!0.5!(8)$); 
    \coordinate (14) at ($(8)!0.5!(2)$); 
    \coordinate (1) at ($(2)!0.5!(3)$); 
    \coordinate (5) at ($(3)!0.5!(4)$); 
    \coordinate (16) at ($(4)!0.5!(10)$); 
    \coordinate (11) at ($(10)!0.5!(9)$); 


    \node at ($(7) + (60:0.3)$) {\color{black}7};
    \node at ($(14) + (0:0.3)$) {\color{black}14};
    \node at ($(1) + (-60:0.3)$) {\color{black}1};
    \node at ($(5) + (-120:0.3)$) {\color{black}5};
    \node at ($(16) + (180:0.3)$) {\color{black}16};
    \node at ($(11) + (120:0.3)$) {\color{black}11};

    \coordinate (15) at (90:1);    
    \node at ($(15) + (90:0.3)$) {\color{black}15};

    \draw[orange]  (16) --  (15);
    \coordinate (17) at ($(16)!0.5!(15)$); 
    \node at ($(17) + (170:0.3)$) {\color{black}17};

    \draw[green]  (15) -- (14);
    \coordinate (13) at ($(15)!0.5!(14)$); 
    \node at ($(13) + (10:0.3)$) {\color{black}13};

    \draw[gray]  (7) -- (5);
    \coordinate (6) at ($(7)!0.5!(5)$); 
    \node at ($(6) + (0:0.3)$) {\color{black}6};

    \draw[brown]  (17) -- (1);
    \coordinate (18) at ($(17)!0.7!(1)$); 
    \node at ($(18) + (190:0.3)$) {\color{black}18};

    \draw[violet] (11) .. controls (100:0.5) .. (13);
    \path (11) .. controls (100:0.5) .. (13) coordinate[pos=0.14] (12);

    \node at ($(12) + (75:0.3)$) {\color{black}12};

    \draw [red] (1) -- (3);
    \draw [red] (1) -- (2);

    \draw [blue] (3) -- (5);
    \draw [blue] (4) -- (5);


    \draw [magenta] (7) -- (9);
    \draw [magenta] (7) -- (8);

    \draw [cyan] (9) -- (11);
    \draw [cyan] (10) -- (11);

    \draw [olive] (4) -- (10);
    \draw [teal] (8) -- (2);

    \draw [ultra thick,gray,dotted] (5) -- (11);
    \draw [ultra thick,gray,dotted] (1) -- (7);

    \fill[green ] (1) circle  (4pt);
    \fill[green] (2) circle  (4pt);
    \fill[red, decorate, decoration=zigzag] (3) circle  (4pt);
    \fill[green] (4) circle  (4pt);
    \fill[green ] (5) circle  (4pt);
    \fill[red, decorate, decoration=zigzag] (6) circle  (4pt);
    \fill[green] (7) circle  (4pt);
    \fill[red, decorate, decoration=zigzag ] (8) circle  (4pt);
    \fill[green] (9) circle  (4pt);
    \fill[red, decorate, decoration=zigzag ] (10) circle (4pt);
    \fill[green] (11) circle (4pt);
    \fill[red, decorate, decoration=zigzag ] (12) circle (4pt);
    \fill[green] (13) circle (4pt);
    \fill[green] (14) circle (4pt);
    \fill[red, decorate, decoration=zigzag ] (15) circle (4pt);
    \fill[green] (16) circle (4pt);
    \fill[green] (17) circle (4pt);
    \fill[red, decorate, decoration=zigzag] (18) circle (4pt);

\end{tikzpicture}
}
&$\qquad$&
\resizebox{.27\textwidth}{!}{
\begin{tikzpicture} [scale=2]
\tikzstyle{every path}=[line width=2pt]

    \coordinate (9) at (90:2);    
    \coordinate (8) at (30:2);    
    \coordinate (2) at (-30:2);   
    \coordinate (3) at (-90:2);   
    \coordinate (4) at (-150:2);  
    \coordinate (10) at (150:2);   

    \node at ($(9) + (90:0.3)$) {\color{black}9};
    \node at ($(8) + (30:0.3)$) {\color{black}8};
    \node at ($(2) + (-30:0.3)$) {\color{black}2};
    \node at ($(3) + (-90:0.3)$) {\color{black}3};
    \node at ($(4) + (-150:0.3)$) {\color{black}4};
    \node at ($(10) + (150:0.3)$) {\color{black}10};

    \coordinate (7) at ($(9)!0.5!(8)$); 
    \coordinate (14) at ($(8)!0.5!(2)$); 
    \coordinate (1) at ($(2)!0.5!(3)$); 
    \coordinate (5) at ($(3)!0.5!(4)$); 
    \coordinate (16) at ($(4)!0.5!(10)$); 
    \coordinate (11) at ($(10)!0.5!(9)$); 


    \node at ($(7) + (60:0.3)$) {\color{black}7};
    \node at ($(14) + (0:0.3)$) {\color{black}14};
    \node at ($(1) + (-60:0.3)$) {\color{black}1};
    \node at ($(5) + (-120:0.3)$) {\color{black}5};
    \node at ($(16) + (180:0.3)$) {\color{black}16};
    \node at ($(11) + (120:0.3)$) {\color{black}11};

    \coordinate (15) at (90:1);    
    \node at ($(15) + (90:0.3)$) {\color{black}15};

    \draw[orange]  (16) --  (15);
    \coordinate (17) at ($(16)!0.5!(15)$); 
    \node at ($(17) + (170:0.3)$) {\color{black}17};

    \draw[green]  (15) -- (14);
    \coordinate (13) at ($(15)!0.5!(14)$); 
    \node at ($(13) + (10:0.3)$) {\color{black}13};

    \draw[gray]  (7) -- (5);
    \coordinate (6) at ($(7)!0.5!(5)$); 
    \node at ($(6) + (0:0.3)$) {\color{black}6};

    \draw[brown]  (17) -- (1);
    \coordinate (18) at ($(17)!0.7!(1)$); 
    \node at ($(18) + (190:0.3)$) {\color{black}18};

    \draw[violet] (11) .. controls (100:0.5) .. (13);
    \path (11) .. controls (100:0.5) .. (13) coordinate[pos=0.14] (12);

    \node at ($(12) + (75:0.3)$) {\color{black}12};

    \draw [red] (1) -- (3);
    \draw [red] (1) -- (2);

    \draw [blue] (3) -- (5);
    \draw [blue] (4) -- (5);


    \draw [magenta] (7) -- (9);
    \draw [magenta] (7) -- (8);

    \draw [cyan] (9) -- (11);
    \draw [cyan] (10) -- (11);

    \draw [olive] (4) -- (10);
    \draw [teal] (8) -- (2);

    \draw [ultra thick,gray,dotted] (5) -- (11);
    \draw [ultra thick,gray,dotted] (1) -- (7);

    \fill[green ] (1) circle  (4pt);
    \fill[green] (2) circle  (4pt);
    \fill[red, decorate, decoration=zigzag] (3) circle  (4pt);
    \fill[green] (4) circle  (4pt);
    \fill[green ] (5) circle  (4pt);
    \fill[red, decorate, decoration=zigzag] (6) circle  (4pt);
    \fill[green] (7) circle  (4pt);
    \fill[green] (8) circle  (4pt);
    \fill[red, decorate, decoration=zigzag ] (9) circle  (4pt);
    \fill[green] (10) circle (4pt);
    \fill[green] (11) circle (4pt);
    \fill[red, decorate, decoration=zigzag] (12) circle (4pt);
    \fill[green] (13) circle (4pt);
    \fill[red, decorate, decoration=zigzag] (14) circle (4pt);
    \fill[green] (15) circle (4pt);
    \fill[red, decorate, decoration=zigzag] (16) circle (4pt);
    \fill[green] (17) circle (4pt);
    \fill[red, decorate, decoration=zigzag] (18) circle (4pt);

\end{tikzpicture}
}
\\
(1)&&(2)&&(3)
\end{tabular}
\end{center}
\caption{\label{2023-e-f-color-tws}
Three two-valued states---numbers 4, 8, and 10---of the M\"obius-Escher-Penrose hypergraph, as listed in Table~\ref{2024-e-t2},
cannot be extended to a 3-coloring.
}
\end{figure}

\section{Loosening tightness}

Some hypergraphs with less interwining contexts than the M\"obius--Escher--Penrose hypergraph
are shown
in Figure~\ref{2023-e-f3}:\\
(a) A variant with two fewer contexts but incorporating an additional intertwining context.\\
(b) A periodic diagram featuring nine cyclically intertwining contexts, with two contexts absent.\\
(c) A periodic diagram with six cyclically intertwining contexts, effectively forming a hexagon.

\begin{figure}
\begin{center}
\begin{tabular}{ccc}
\resizebox{.3\textwidth}{!}{\begin{tikzpicture}  [scale=1] 
\tikzstyle{every path}=[line width=2pt]

\tikzmath{\a = 1; \b = 2; \c = 0;}

\newdimen\ms
\ms=0.05cm

\tikzstyle{c3}=[circle,inner sep={\ms/8},minimum size=6*\ms]
\tikzstyle{c2}=[circle,inner sep={\ms/8},minimum size=4*\ms]
\tikzstyle{c1}=[circle,inner sep={\ms/8},minimum size=0.8*\ms]

\coordinate (1) at ({2*( \a +  \a + \b ) + 2*\c},{( \a +  \a + \b ) + 2*\c});
\coordinate (3) at ({( \a +  \a + \b )},0);
\coordinate (2) at ($(1)!0.5!(3)$);
\coordinate (5) at ({0-2*\c},{( \a +  \a + \b )+2*\c});
\coordinate (4) at ($(3)!0.5!(5)$);
\coordinate (13) at ({2*( \a +  \a + \b )-\a+\c},{( \a +  \a + \b )+\c});
\coordinate (15) at ({( \a +  \a + \b )},\a);
\coordinate (14) at ($(13)!0.5!(15)$);
\coordinate (17) at ({\a-\c},{( \a +  \a + \b )+\c});
\coordinate (16) at ($(15)!0.5!(17)$);
\coordinate (7) at ({2*( \a +  \a + \b )-2*\a},{( \a +  \a + \b )});
\coordinate (9) at ({( \a +  \a + \b )},{2*\a});
\coordinate (8) at ($(7)!0.5!(9)$);
\coordinate (11) at ({2*\a},{( \a +  \a + \b )});
\coordinate (10) at ($(9)!0.5!(11)$);

 \coordinate (12) at ($(11)!0.5!(13)$);
 \coordinate (18) at ($(17)!0.5!(1)$);

\draw [red] (1) -- (3);
\draw [blue] (3) -- (5);

\draw [green] (13) -- (15);
\draw [orange] (15) -- (17);

\draw [magenta] (7) -- (9);
\draw [cyan] (9) -- (11);

\draw [gray] (5) .. controls (\a,{1.6*( \a +  \a + \b )}) and ({ 2*( \a +  \a + \b ) - 2} ,{1.6*( \a +  \a + \b )}) .. (7)
                coordinate [pos=0.3] (6);

\draw [brown] (17) .. controls ({ ( \a +  \a + \b ) - 2},{1.5*( \a +  \a + \b )}) and ({  ( \a +  \a + \b ) + 2} ,{1.5*( \a +  \a + \b )}) .. (1)
                coordinate [pos=0.223] (18);

\draw [violet] (11) .. controls ({ ( \a +  \a + \b ) - 2},{1.3*( \a +  \a + \b )}) and ({ ( \a +  \a + \b ) + 2} ,{1.3*( \a +  \a + \b )}) .. (13)
                coordinate [pos=0.2] (12);

\draw [teal] (6) -- (12);

\draw (1) coordinate[c3,fill=red,label={below right: $1$}];
\draw (2) coordinate[c3,fill=red,label={below right: $2$}];
\draw (3) coordinate[c3,fill=red,label={below: $3$}];
\draw (4) coordinate[c3,fill=blue,label={below: $4$}];
\draw (5) coordinate[c3,fill=blue,label={below: $5$}];
\draw (6) coordinate[c3,fill=gray,label={above left: $6$}];
\draw (7) coordinate[c3,fill=magenta,label={below: $7$}];
\draw (8) coordinate[c3,fill=magenta,label={below: $8$}];
\draw (9) coordinate[c3,fill=cyan,label={below: $9$}];
\draw (10) coordinate[c3,fill=cyan,label={below: $10$}];
\draw (11) coordinate[c3,fill=cyan,label={below: $11$}];
\draw (12) coordinate[c3,fill=violet,label={below: $12$}];
\draw (13) coordinate[c3,fill=green,label={below: $13$}];
\draw (14) coordinate[c3,fill=green,label={below: $14$}];
\draw (15) coordinate[c3,fill=orange,label={below: $15$}];
\draw (16) coordinate[c3,fill=orange,label={below: $16$}];
\draw (17) coordinate[c3,fill=orange,label={below: $17$}];
\draw (18) coordinate[c3,fill=brown,label={left: $18$}];

\draw (9) coordinate[c2,fill=magenta];
\draw (15) coordinate[c2,fill=green];
\draw (3) coordinate[c2,fill=blue];

\draw (5) coordinate[c2,fill=gray];
\draw (17) coordinate[c2,fill=brown];
\draw (11) coordinate[c2,fill=violet];

\draw (7) coordinate[c2,fill=gray];
\draw (1) coordinate[c2,fill=brown];
\draw (13) coordinate[c2,fill=violet];

\draw (12) coordinate[c2,fill=teal];
\draw (18) coordinate[c2,fill=teal];
\draw (6) coordinate[c2,fill=teal];

\end{tikzpicture}}
&
\resizebox{.3\textwidth}{!}{\begin{tikzpicture}  [scale=1] 
\tikzstyle{every path}=[line width=2pt]

\tikzmath{\a = 1; \b = 2; \c = 0;}

\newdimen\ms
\ms=0.05cm

\tikzstyle{c3}=[circle,inner sep={\ms/8},minimum size=6*\ms]
\tikzstyle{c2}=[circle,inner sep={\ms/8},minimum size=4*\ms]
\tikzstyle{c1}=[circle,inner sep={\ms/8},minimum size=0.8*\ms]

\coordinate (1) at ({2*( \a +  \a + \b ) + 2*\c},{( \a +  \a + \b ) + 2*\c});
\coordinate (3) at ({( \a +  \a + \b )},0);
\coordinate (2) at ($(1)!0.5!(3)$);
\coordinate (5) at ({0-2*\c},{( \a +  \a + \b )+2*\c});
\coordinate (4) at ($(3)!0.5!(5)$);
\coordinate (13) at ({2*( \a +  \a + \b )-\a+\c},{( \a +  \a + \b )+\c});
\coordinate (15) at ({( \a +  \a + \b )},\a);
\coordinate (14) at ($(13)!0.5!(15)$);
\coordinate (17) at ({\a-\c},{( \a +  \a + \b )+\c});
\coordinate (16) at ($(15)!0.5!(17)$);
\coordinate (7) at ({2*( \a +  \a + \b )-2*\a},{( \a +  \a + \b )});
\coordinate (9) at ({( \a +  \a + \b )},{2*\a});
\coordinate (8) at ($(7)!0.5!(9)$);
\coordinate (11) at ({2*\a},{( \a +  \a + \b )});
\coordinate (10) at ($(9)!0.5!(11)$);

 \coordinate (12) at ($(11)!0.5!(13)$);
 \coordinate (18) at ($(17)!0.5!(1)$);

\draw [red] (1) -- (3);
\draw [blue] (3) -- (5);

\draw [green] (13) -- (15);
\draw [orange] (15) -- (17);

\draw [magenta] (7) -- (9);
\draw [cyan] (9) -- (11);

\draw [gray] (5) .. controls (\a,{1.6*( \a +  \a + \b )}) and ({ 2*( \a +  \a + \b ) - 2} ,{1.6*( \a +  \a + \b )}) .. (7)
                coordinate [pos=0.3] (6);

\draw [brown] (17) .. controls ({ ( \a +  \a + \b ) - 2},{1.5*( \a +  \a + \b )}) and ({  ( \a +  \a + \b ) + 2} ,{1.5*( \a +  \a + \b )}) .. (1)
                coordinate [pos=0.8] (18);

\draw [violet] (11) .. controls ({ ( \a +  \a + \b ) - 2},{1.3*( \a +  \a + \b )}) and ({ ( \a +  \a + \b ) + 2} ,{1.3*( \a +  \a + \b )}) .. (13)
                coordinate [pos=0.2] (12);

\draw (1) coordinate[c3,fill=red,label={below right: $1$}];
\draw (2) coordinate[c3,fill=red,label={below right: $2$}];
\draw (3) coordinate[c3,fill=red,label={below: $3$}];
\draw (4) coordinate[c3,fill=blue,label={below: $4$}];
\draw (5) coordinate[c3,fill=blue,label={below: $5$}];
\draw (6) coordinate[c3,fill=gray,label={above left: $6$}];
\draw (7) coordinate[c3,fill=magenta,label={below: $7$}];
\draw (8) coordinate[c3,fill=magenta,label={below: $8$}];
\draw (9) coordinate[c3,fill=cyan,label={below: $9$}];
\draw (10) coordinate[c3,fill=cyan,label={below: $10$}];
\draw (11) coordinate[c3,fill=cyan,label={below: $11$}];
\draw (12) coordinate[c3,fill=violet,label={below: $12$}];
\draw (13) coordinate[c3,fill=green,label={below: $13$}];
\draw (14) coordinate[c3,fill=green,label={below: $14$}];
\draw (15) coordinate[c3,fill=orange,label={below: $15$}];
\draw (16) coordinate[c3,fill=orange,label={below: $16$}];
\draw (17) coordinate[c3,fill=orange,label={below: $17$}];
\draw (18) coordinate[c3,fill=brown,label={above: $18$}];

\draw (9) coordinate[c2,fill=magenta];
\draw (15) coordinate[c2,fill=green];
\draw (3) coordinate[c2,fill=blue];

\draw (5) coordinate[c2,fill=gray];
\draw (17) coordinate[c2,fill=brown];
\draw (11) coordinate[c2,fill=violet];

\draw (7) coordinate[c2,fill=gray];
\draw (1) coordinate[c2,fill=brown];
\draw (13) coordinate[c2,fill=violet];

\end{tikzpicture}}
&
\resizebox{.3\textwidth}{!}{\begin{tikzpicture}  [scale=1] 
\tikzstyle{every path}=[line width=2pt]

\tikzmath{\a = 1; \b = 2; \c = 0;}

\newdimen\ms
\ms=0.05cm

\tikzstyle{c3}=[circle,inner sep={\ms/8},minimum size=6*\ms]
\tikzstyle{c2}=[circle,inner sep={\ms/8},minimum size=4*\ms]
\tikzstyle{c1}=[circle,inner sep={\ms/8},minimum size=0.8*\ms]

\coordinate (1) at ({2*( \a +  \a + \b ) + 2*\c},{( \a +  \a + \b ) + 2*\c});
\coordinate (3) at ({( \a +  \a + \b )},0);
\coordinate (2) at ($(1)!0.5!(3)$);
\coordinate (5) at ({0-2*\c},{( \a +  \a + \b )+2*\c});
\coordinate (4) at ($(3)!0.5!(5)$);
\coordinate (13) at ({2*( \a +  \a + \b )-\a+\c},{( \a +  \a + \b )+\c});
\coordinate (15) at ({( \a +  \a + \b )},\a);
\coordinate (14) at ($(13)!0.5!(15)$);
\coordinate (17) at ({\a-\c},{( \a +  \a + \b )+\c});
\coordinate (16) at ($(15)!0.5!(17)$);
\coordinate (7) at ({2*( \a +  \a + \b )-2*\a},{( \a +  \a + \b )});
\coordinate (9) at ({( \a +  \a + \b )},{2*\a});
\coordinate (8) at ($(7)!0.5!(9)$);
\coordinate (11) at ({2*\a},{( \a +  \a + \b )});
\coordinate (10) at ($(9)!0.5!(11)$);

 \coordinate (12) at ($(11)!0.5!(13)$);
 \coordinate (18) at ($(17)!0.5!(1)$);

\draw [red] (1) -- (3);
\draw [blue] (3) -- (5);

\draw [magenta] (7) -- (9);
\draw [cyan] (9) -- (11);

\draw [gray] (5) .. controls (\a,{1.6*( \a +  \a + \b )}) and ({ 2*( \a +  \a + \b ) - 2} ,{1.6*( \a +  \a + \b )}) .. (7)
                coordinate [pos=0.3] (6);

\draw [violet] (11) .. controls ({ ( \a +  \a + \b ) - 2},{1.6*( \a +  \a + \b )}) and ({ ( \a +  \a + \b ) + 2} ,{1.6*( \a +  \a + \b )}) .. (1)
                coordinate [pos=0.7] (12);

\draw (1) coordinate[c3,fill=red,label={below right: $1$}];
\draw (2) coordinate[c3,fill=red,label={below right: $2$}];
\draw (3) coordinate[c3,fill=red,label={below: $3$}];
\draw (4) coordinate[c3,fill=blue,label={below: $4$}];
\draw (5) coordinate[c3,fill=blue,label={below: $5$}];
\draw (6) coordinate[c3,fill=gray,label={above left: $6$}];
\draw (7) coordinate[c3,fill=magenta,label={below: $7$}];
\draw (8) coordinate[c3,fill=magenta,label={below: $8$}];
\draw (9) coordinate[c3,fill=cyan,label={below: $9$}];
\draw (10) coordinate[c3,fill=cyan,label={below: $10$}];
\draw (11) coordinate[c3,fill=cyan,label={below: $11$}];
\draw (12) coordinate[c3,fill=violet,label={above: $12$}];

\draw (9) coordinate[c2,fill=magenta];
\draw (3) coordinate[c2,fill=blue];

\draw (5) coordinate[c2,fill=gray];
\draw (11) coordinate[c2,fill=violet];

\draw (7) coordinate[c2,fill=gray];

\end{tikzpicture}}
\\
(a)&(b)&(c)
\end{tabular}
\end{center}
\caption{\label{2023-e-f3}
Some hypergraphs with less intertwining contexts than the M\"obius--Escher--Penrose hypergraph:
(a) A variant with two fewer contexts but incorporating an additional intertwining context.
(b) A periodic diagram featuring nine cyclically intertwining contexts, with two contexts absent.
(c) A periodic diagram with six cyclically intertwining contexts, effectively forming a hexagon.
}
\end{figure}

We have found the following coordinatization of the hypergraph from Figure~\ref{2023-e-f3}(a):

We start from the context $\{1,17,18\}$, chosen as
\begin{align*}
\vert v_1 \rangle & = \vert y \rangle = \begin{pmatrix}
0,1,0\end{pmatrix}^\intercal\,, \\
\vert v_{17} \rangle &= \begin{pmatrix}1/\sqrt{3}, 0, -\sqrt{2/3}\end{pmatrix}^\intercal\,, \\
\vert v_{18} \rangle &= \begin{pmatrix}\sqrt{2/3}, 0, 1/\sqrt{3}\end{pmatrix}^\intercal\,.
\end{align*}
Contexts $\{7,5,6\}$ and $\{13,11,12\}$
(in this order of elements) are obtained by the rotation of the context $\{1,17,18\}$ around the $z$-axis by $(2/3)\pi$ and $(4/3)\pi$, respectively.
Trough these rotations, the vectors $v_{18}$,
$v_6=\begin{pmatrix}-1/\sqrt{6},1/\sqrt{2},1/\sqrt{3}\end{pmatrix}^\intercal$,
and
$v_{12}=\begin{pmatrix}-1/\sqrt{6},-1/\sqrt{2},1/\sqrt{3}\end{pmatrix}^\intercal$
form an orthonormal basis, representing the context $\{6,12,18\}$.
The remaining vectors are now uniquely determined (up to the orientation, which is unimportant) and can be computed by cross products.
Vector $v_3$ is a~unit vector in the direction $v_1 \times v_5$ (these are not orthogonal) and
$v_2 = v_1 \times v_3$,
$v_4 = v_5 \times v_3$.
Analogously,
$v_9,v_8,v_{10}$ are constructed from $v_7,v_{11}$
and
$v_{15},v_{14},v_{16}$ from $v_{13},v_{17}$.
We checked that all the $18$ vectors are distinct and satisfy no other orthogonality relations than those denoted in Figure~\ref{2023-e-f3}(a).

\section{Summary}

In this paper, we have introduced and explored the quantum M\"obius--Escher--Penrose hypergraph,
a quantum analog inspired by paradoxical drawings and concepts such as the M\"obius strip and Penrose's `impossible object'.
We have provided a detailed construction of this hypergraph using orthogonal representations
where edges correspond to orthonormal bases in Hilbert space, ensuring a faithful embedding of the graph into a quantum mechanical framework.

The hypergraph can also be realized quasiclassically
by computing two-valued states and creating a partition logic,
establishing an embedding into a Boolean algebra.
This dual approach highlights the distinction between classical and quantum mechanical
embeddings and the emergence of contextuality.

Contextuality, a hallmark of quantum mechanics,
is evidenced by the violation of exclusivity and completeness in certain contexts,
demonstrated both through the inspection of two-valued states and proof by contradiction.
We have quantified these violations using classical probability bounds and show maximal
violations with quantum probabilities for specific states.

Further, we have explored violations of Boole's `conditions of possible experience' through correlation polytopes,
translating binary value assignments into quantum mechanical expectations.
The study reveals multiple hull inequalities,
with the longest summations showing significant quantum violations, thus illustrating the inherent contextuality of the hypergraph.

Our findings contribute to a deeper understanding of quantum contextuality,
extending beyond traditional Hardy-type paradoxes and state-independent proofs,
and underscore the intricate relationship between classical logic structures and quantum mechanical systems.

\begin{acknowledgments}
We are grateful to Josef Tkadlec for providing a {\em Pascal} program that computes and analyses the set of two-valued states of collections of contexts.
We are also grateful to  Norman D. Megill and Mladen Pavi{\v{c}}i{\'{c}} for providing a {\em C++} program that heuristically computes the faithful orthogonal representations of hypergraphs written in MMP format, given possible vector components.

MN was supported by the CTU institutional support (Future Fund).
KS was funded in whole or in part by the Austrian Science Fund (FWF) [Grant DOI:10.55776/I4579].

The authors declare no conflict of interest.
\end{acknowledgments}

\bibliography{svozil}

\begin{thebibliography}{35}%
\makeatletter
\providecommand \@ifxundefined [1]{%
 \@ifx{#1\undefined}
}%
\providecommand \@ifnum [1]{%
 \ifnum #1\expandafter \@firstoftwo
 \else \expandafter \@secondoftwo
 \fi
}%
\providecommand \@ifx [1]{%
 \ifx #1\expandafter \@firstoftwo
 \else \expandafter \@secondoftwo
 \fi
}%
\providecommand \natexlab [1]{#1}%
\providecommand \enquote  [1]{``#1''}%
\providecommand \bibnamefont  [1]{#1}%
\providecommand \bibfnamefont [1]{#1}%
\providecommand \citenamefont [1]{#1}%
\providecommand \href@noop [0]{\@secondoftwo}%
\providecommand \href [0]{\begingroup \@sanitize@url \@href}%
\providecommand \@href[1]{\@@startlink{#1}\@@href}%
\providecommand \@@href[1]{\endgroup#1\@@endlink}%
\providecommand \@sanitize@url [0]{\catcode `\\12\catcode `\$12\catcode
  `\&12\catcode `\#12\catcode `\^12\catcode `\_12\catcode `\%12\relax}%
\providecommand \@@startlink[1]{}%
\providecommand \@@endlink[0]{}%
\providecommand \url  [0]{\begingroup\@sanitize@url \@url }%
\providecommand \@url [1]{\endgroup\@href {#1}{\urlprefix }}%
\providecommand \urlprefix  [0]{URL }%
\providecommand \Eprint [0]{\href }%
\providecommand \doibase [0]{https://doi.org/}%
\providecommand \selectlanguage [0]{\@gobble}%
\providecommand \bibinfo  [0]{\@secondoftwo}%
\providecommand \bibfield  [0]{\@secondoftwo}%
\providecommand \translation [1]{[#1]}%
\providecommand \BibitemOpen [0]{}%
\providecommand \bibitemStop [0]{}%
\providecommand \bibitemNoStop [0]{.\EOS\space}%
\providecommand \EOS [0]{\spacefactor3000\relax}%
\providecommand \BibitemShut  [1]{\csname bibitem#1\endcsname}%
\let\auto@bib@innerbib\@empty
\bibitem [{\citenamefont {Penrose}\ and\ \citenamefont
  {Penrose}(1958)}]{PENROSE_1958}%
  \BibitemOpen
  \bibfield  {author} {\bibinfo {author} {\bibfnamefont {L.~S.}\ \bibnamefont
  {Penrose}}\ and\ \bibinfo {author} {\bibfnamefont {R.}~\bibnamefont
  {Penrose}},\ }\bibfield  {title} {\bibinfo {title} {Impossible objects: {A}
  special type of visual illusion},\ }\href
  {https://doi.org/10.1111/j.2044-8295.1958.tb00634.x} {\bibfield  {journal}
  {\bibinfo  {journal} {British Journal of Psychology}\ }\textbf {\bibinfo
  {volume} {49}},\ \bibinfo {pages} {31} (\bibinfo {year} {1958})}\BibitemShut
  {NoStop}%
\bibitem [{\citenamefont {Escher}(1954)}]{Escher1954}%
  \BibitemOpen
  \bibfield  {author} {\bibinfo {author} {\bibfnamefont {M.~C.}\ \bibnamefont
  {Escher}},\ }\href@noop {} {\emph {\bibinfo {title} {M. C. Escher: Catalogus
  118}}}\ (\bibinfo  {publisher} {Stedelijk Museum},\ \bibinfo {address}
  {Amsterdam, Netherlands},\ \bibinfo {year} {1954})\ \bibinfo {note}
  {catalogue for the M. C. Escher Exhibition at the Stedelijk Museum. The
  Organizing Committee of the International Congress of Mathematicians 1954
  (Sept. 2nd-9th) at Amsterdam, Summer 1954. Book Design by Willem
  Sandberg}\BibitemShut {NoStop}%
\bibitem [{\citenamefont {Cartwright}\ and\ \citenamefont
  {Gonz\'alez}(2016)}]{cartwright-2016}%
  \BibitemOpen
  \bibfield  {author} {\bibinfo {author} {\bibfnamefont {J.~H.~E.}\
  \bibnamefont {Cartwright}}\ and\ \bibinfo {author} {\bibfnamefont {D.~L.}\
  \bibnamefont {Gonz\'alez}},\ }\bibfield  {title} {\bibinfo {title}
  {{M}\"obius strips before {M}\"obius: {T}opological hints in ancient
  representations},\ }\href {https://doi.org/10.1007/s00283-016-9631-8}
  {\bibfield  {journal} {\bibinfo  {journal} {The Mathematical Intelligencer}\
  }\textbf {\bibinfo {volume} {38}},\ \bibinfo {pages} {69} (\bibinfo {year}
  {2016})}\BibitemShut {NoStop}%
\bibitem [{\citenamefont {Navara}\ and\ \citenamefont
  {Svozil}(2024)}]{2023-navara-svozil}%
  \BibitemOpen
  \bibfield  {author} {\bibinfo {author} {\bibfnamefont {M.}~\bibnamefont
  {Navara}}\ and\ \bibinfo {author} {\bibfnamefont {K.}~\bibnamefont
  {Svozil}},\ }\bibfield  {title} {\bibinfo {title} {Form of contextuality
  predicting probabilistic equivalence between two sets of three mutually
  noncommuting observables},\ }\href
  {https://doi.org/10.1103/PhysRevA.109.022222} {\bibfield  {journal} {\bibinfo
   {journal} {Physical Review A}\ }\textbf {\bibinfo {volume} {109}},\ \bibinfo
  {pages} {022222} (\bibinfo {year} {2024})},\ \Eprint
  {https://arxiv.org/abs/arXiv:2309.13091} {arXiv:2309.13091} \BibitemShut
  {NoStop}%
\bibitem [{\citenamefont {Kochen}\ and\ \citenamefont
  {Specker}(1967)}]{kochen1}%
  \BibitemOpen
  \bibfield  {author} {\bibinfo {author} {\bibfnamefont {S.}~\bibnamefont
  {Kochen}}\ and\ \bibinfo {author} {\bibfnamefont {E.~P.}\ \bibnamefont
  {Specker}},\ }\bibfield  {title} {\bibinfo {title} {The problem of hidden
  variables in quantum mechanics},\ }\href
  {https://doi.org/10.1512/iumj.1968.17.17004} {\bibfield  {journal} {\bibinfo
  {journal} {Journal of Mathematics and Mechanics (now Indiana University
  Mathematics Journal)}\ }\textbf {\bibinfo {volume} {17}},\ \bibinfo {pages}
  {59} (\bibinfo {year} {1967})}\BibitemShut {NoStop}%
\bibitem [{\citenamefont {Lov\'asz}(1979)}]{lovasz-79}%
  \BibitemOpen
  \bibfield  {author} {\bibinfo {author} {\bibfnamefont {L.}~\bibnamefont
  {Lov\'asz}},\ }\bibfield  {title} {\bibinfo {title} {On the {S}hannon
  capacity of a graph},\ }\href {https://doi.org/10.1109/TIT.1979.1055985}
  {\bibfield  {journal} {\bibinfo  {journal} {IEEE Transactions on Information
  Theory}\ }\textbf {\bibinfo {volume} {25}},\ \bibinfo {pages} {1} (\bibinfo
  {year} {1979})}\BibitemShut {NoStop}%
\bibitem [{\citenamefont {Gr{\"o}tschel}\ \emph {et~al.}(1986)\citenamefont
  {Gr{\"o}tschel}, \citenamefont {Lov{\'a}sz},\ and\ \citenamefont
  {Schrijver}}]{GroetschelLovaszSchrijver1986}%
  \BibitemOpen
  \bibfield  {author} {\bibinfo {author} {\bibfnamefont {M.}~\bibnamefont
  {Gr{\"o}tschel}}, \bibinfo {author} {\bibfnamefont {L.}~\bibnamefont
  {Lov{\'a}sz}},\ and\ \bibinfo {author} {\bibfnamefont {A.}~\bibnamefont
  {Schrijver}},\ }\bibfield  {title} {\bibinfo {title} {Relaxations of vertex
  packing},\ }\href {https://doi.org/10.1016/0095-8956(86)90087-0} {\bibfield
  {journal} {\bibinfo  {journal} {Journal of Combinatorial Theory, Series B}\
  }\textbf {\bibinfo {volume} {40}},\ \bibinfo {pages} {330} (\bibinfo {year}
  {1986})}\BibitemShut {NoStop}%
\bibitem [{\citenamefont {Sol\'is-Encina}\ and\ \citenamefont
  {Portillo}(2015)}]{Portillo-2015}%
  \BibitemOpen
  \bibfield  {author} {\bibinfo {author} {\bibfnamefont {A.}~\bibnamefont
  {Sol\'is-Encina}}\ and\ \bibinfo {author} {\bibfnamefont {J.~R.}\
  \bibnamefont {Portillo}},\ }\href {https://doi.org/10.48550/arXiv.1504.03662}
  {\bibinfo {title} {Orthogonal representation of graphs}} (\bibinfo {year}
  {2015}),\ \Eprint {https://arxiv.org/abs/arXiv:1504.03662} {arXiv:1504.03662}
  \BibitemShut {NoStop}%
\bibitem [{\citenamefont {Navara}\ and\ \citenamefont
  {Rogalewicz}(1991)}]{nav:91}%
  \BibitemOpen
  \bibfield  {author} {\bibinfo {author} {\bibfnamefont {M.}~\bibnamefont
  {Navara}}\ and\ \bibinfo {author} {\bibfnamefont {V.}~\bibnamefont
  {Rogalewicz}},\ }\bibfield  {title} {\bibinfo {title} {The pasting
  constructions for orthomodular posets},\ }\href
  {https://doi.org/10.1002/mana.19911540113} {\bibfield  {journal} {\bibinfo
  {journal} {Mathematische Nachrichten}\ }\textbf {\bibinfo {volume} {154}},\
  \bibinfo {pages} {157} (\bibinfo {year} {1991})}\BibitemShut {NoStop}%
\bibitem [{\citenamefont {Svozil}(2005)}]{svozil-2001-eua}%
  \BibitemOpen
  \bibfield  {author} {\bibinfo {author} {\bibfnamefont {K.}~\bibnamefont
  {Svozil}},\ }\bibfield  {title} {\bibinfo {title} {Logical equivalence
  between generalized urn models and finite automata},\ }\href
  {https://doi.org/10.1007/s10773-005-7052-0} {\bibfield  {journal} {\bibinfo
  {journal} {International Journal of Theoretical Physics}\ }\textbf {\bibinfo
  {volume} {44}},\ \bibinfo {pages} {745} (\bibinfo {year} {2005})},\ \Eprint
  {https://arxiv.org/abs/arXiv:quant-ph/0209136} {arXiv:quant-ph/0209136}
  \BibitemShut {NoStop}%
\bibitem [{\citenamefont {Zierler}\ and\ \citenamefont
  {Schlessinger}(1965)}]{ZirlSchl-65}%
  \BibitemOpen
  \bibfield  {author} {\bibinfo {author} {\bibfnamefont {N.}~\bibnamefont
  {Zierler}}\ and\ \bibinfo {author} {\bibfnamefont {M.}~\bibnamefont
  {Schlessinger}},\ }\bibfield  {title} {\bibinfo {title} {Boolean embeddings
  of orthomodular sets and quantum logic},\ }\href
  {https://doi.org/10.1215/S0012-7094-65-03224-2} {\bibfield  {journal}
  {\bibinfo  {journal} {Duke Mathematical Journal}\ }\textbf {\bibinfo {volume}
  {32}},\ \bibinfo {pages} {251} (\bibinfo {year} {1965})},\ \bibinfo {note}
  {reprinted in Ref.~\cite{Zierler1975}}\BibitemShut {NoStop}%
\bibitem [{\citenamefont {Travis}(1962)}]{travis-mt-62}%
  \BibitemOpen
  \bibfield  {author} {\bibinfo {author} {\bibfnamefont {R.~D.}\ \bibnamefont
  {Travis}},\ }\emph {\bibinfo {title} {The Logic of a Physical Theory}},\
  \href@noop {} {Master's thesis},\ \bibinfo  {school} {Wayne State
  University}, \bibinfo {address} {Detroit, Michigan, USA} (\bibinfo {year}
  {1962}),\ \bibinfo {note} {{M}aster's {T}hesis under the supervision of David
  J. Foulis}\BibitemShut {NoStop}%
\bibitem [{\citenamefont {Greechie}(1966)}]{greechie-66-PhD}%
  \BibitemOpen
  \bibfield  {author} {\bibinfo {author} {\bibfnamefont {R.~J.}\ \bibnamefont
  {Greechie}},\ }\emph {\bibinfo {title} {Orthomodular Lattices}},\ \href
  {https://ufdc.ufl.edu/UF00097858/00001/pdf} {Ph.D. thesis},\ \bibinfo
  {school} {University of Florida}, \bibinfo {address} {Florida, USA} (\bibinfo
  {year} {1966})\BibitemShut {NoStop}%
\bibitem [{\citenamefont {Cabello}\ \emph {et~al.}(2018)\citenamefont
  {Cabello}, \citenamefont {Portillo}, \citenamefont {Sol\'{i}s},\ and\
  \citenamefont {Svozil}}]{2018-minimalYIYS}%
  \BibitemOpen
  \bibfield  {author} {\bibinfo {author} {\bibfnamefont {A.}~\bibnamefont
  {Cabello}}, \bibinfo {author} {\bibfnamefont {J.~R.}\ \bibnamefont
  {Portillo}}, \bibinfo {author} {\bibfnamefont {A.}~\bibnamefont
  {Sol\'{i}s}},\ and\ \bibinfo {author} {\bibfnamefont {K.}~\bibnamefont
  {Svozil}},\ }\bibfield  {title} {\bibinfo {title} {Minimal true-implies-false
  and true-implies-true sets of propositions in noncontextual hidden-variable
  theories},\ }\href {https://doi.org/10.1103/PhysRevA.98.012106} {\bibfield
  {journal} {\bibinfo  {journal} {Physical Review A}\ }\textbf {\bibinfo
  {volume} {98}},\ \bibinfo {pages} {012106} (\bibinfo {year} {2018})},\
  \Eprint {https://arxiv.org/abs/arXiv:1805.00796} {arXiv:1805.00796}
  \BibitemShut {NoStop}%
\bibitem [{\citenamefont {Kochen}\ and\ \citenamefont
  {Specker}(1965)}]{kochen2}%
  \BibitemOpen
  \bibfield  {author} {\bibinfo {author} {\bibfnamefont {S.}~\bibnamefont
  {Kochen}}\ and\ \bibinfo {author} {\bibfnamefont {E.~P.}\ \bibnamefont
  {Specker}},\ }\bibfield  {title} {\bibinfo {title} {Logical structures
  arising in quantum theory},\ }in\ \href
  {https://doi.org/978-3-0348-9259-9_19} {\emph {\bibinfo {booktitle} {The
  Theory of Models, {P}roceedings of the 1963 International Symposium at
  {B}erkeley}}},\ \bibinfo {editor} {edited by\ \bibinfo {editor}
  {\bibfnamefont {J.~W.}\ \bibnamefont {Addison}}, \bibinfo {editor}
  {\bibfnamefont {L.}~\bibnamefont {Henkin}},\ and\ \bibinfo {editor}
  {\bibfnamefont {A.}~\bibnamefont {Tarski}}}\ (\bibinfo  {publisher} {North
  Holland},\ \bibinfo {address} {Amsterdam, New York, Oxford},\ \bibinfo {year}
  {1965})\ pp.\ \bibinfo {pages} {177--189}\BibitemShut {NoStop}%
\bibitem [{\citenamefont {Specker}(1990)}]{specker-ges}%
  \BibitemOpen
  \bibfield  {author} {\bibinfo {author} {\bibfnamefont {E.}~\bibnamefont
  {Specker}},\ }\href {https://doi.org/10.1007/978-3-0348-9259-9} {\emph
  {\bibinfo {title} {Selecta}}}\ (\bibinfo  {publisher} {Birkh{\"{a}}user
  Verlag},\ \bibinfo {address} {Basel},\ \bibinfo {year} {1990})\BibitemShut
  {NoStop}%
\bibitem [{\citenamefont {Cabello}\ \emph {et~al.}(2013)\citenamefont
  {Cabello}, \citenamefont {Badziag}, \citenamefont {Terra~Cunha},\ and\
  \citenamefont {Bourennane}}]{Cabello-2013-Hardylike}%
  \BibitemOpen
  \bibfield  {author} {\bibinfo {author} {\bibfnamefont {A.}~\bibnamefont
  {Cabello}}, \bibinfo {author} {\bibfnamefont {P.}~\bibnamefont {Badziag}},
  \bibinfo {author} {\bibfnamefont {M.}~\bibnamefont {Terra~Cunha}},\ and\
  \bibinfo {author} {\bibfnamefont {M.}~\bibnamefont {Bourennane}},\ }\bibfield
   {title} {\bibinfo {title} {Simple {H}ardy-like proof of quantum
  contextuality},\ }\href {https://doi.org/10.1103/PhysRevLett.111.180404}
  {\bibfield  {journal} {\bibinfo  {journal} {Physical Review Letters}\
  }\textbf {\bibinfo {volume} {111}},\ \bibinfo {pages} {180404} (\bibinfo
  {year} {2013})}\BibitemShut {NoStop}%
\bibitem [{\citenamefont {Yu}\ and\ \citenamefont {Oh}(2012)}]{Yu-2012}%
  \BibitemOpen
  \bibfield  {author} {\bibinfo {author} {\bibfnamefont {S.}~\bibnamefont
  {Yu}}\ and\ \bibinfo {author} {\bibfnamefont {C.~H.}\ \bibnamefont {Oh}},\
  }\bibfield  {title} {\bibinfo {title} {State-independent proof of
  {K}ochen-{S}pecker theorem with 13 rays},\ }\href
  {https://doi.org/10.1103/PhysRevLett.108.030402} {\bibfield  {journal}
  {\bibinfo  {journal} {Physical Review Letters}\ }\textbf {\bibinfo {volume}
  {108}},\ \bibinfo {pages} {030402} (\bibinfo {year} {2012})},\ \Eprint
  {https://arxiv.org/abs/arXiv:1109.4396} {arXiv:1109.4396} \BibitemShut
  {NoStop}%
\bibitem [{\citenamefont {Budroni}\ \emph {et~al.}(2022)\citenamefont
  {Budroni}, \citenamefont {Cabello}, \citenamefont {G\"uhne}, \citenamefont
  {Kleinmann},\ and\ \citenamefont {Larsson}}]{cabello2021contextuality}%
  \BibitemOpen
  \bibfield  {author} {\bibinfo {author} {\bibfnamefont {C.}~\bibnamefont
  {Budroni}}, \bibinfo {author} {\bibfnamefont {A.}~\bibnamefont {Cabello}},
  \bibinfo {author} {\bibfnamefont {O.}~\bibnamefont {G\"uhne}}, \bibinfo
  {author} {\bibfnamefont {M.}~\bibnamefont {Kleinmann}},\ and\ \bibinfo
  {author} {\bibfnamefont {J.-A.}\ \bibnamefont {Larsson}},\ }\bibfield
  {title} {\bibinfo {title} {{K}ochen-{S}pecker contextuality},\ }\href
  {https://doi.org/10.1103/revmodphys.94.045007} {\bibfield  {journal}
  {\bibinfo  {journal} {Reviews of Modern Physics}\ }\textbf {\bibinfo {volume}
  {94}},\ \bibinfo {pages} {045007} (\bibinfo {year} {2022})},\ \Eprint
  {https://arxiv.org/abs/arXiv:2102.13036} {arXiv:2102.13036} \BibitemShut
  {NoStop}%
\bibitem [{\citenamefont {Boole}(1862)}]{Boole-62}%
  \BibitemOpen
  \bibfield  {author} {\bibinfo {author} {\bibfnamefont {G.}~\bibnamefont
  {Boole}},\ }\bibfield  {title} {\bibinfo {title} {On the theory of
  probabilities},\ }\href {https://doi.org/10.1098/rstl.1862.0015} {\bibfield
  {journal} {\bibinfo  {journal} {Philosophical Transactions of the Royal
  Society of London}\ }\textbf {\bibinfo {volume} {152}},\ \bibinfo {pages}
  {225} (\bibinfo {year} {1862})}\BibitemShut {NoStop}%
\bibitem [{\citenamefont {Svozil}(2022)}]{svozil-2021-hh}%
  \BibitemOpen
  \bibfield  {author} {\bibinfo {author} {\bibfnamefont {K.}~\bibnamefont
  {Svozil}},\ }\bibfield  {title} {\bibinfo {title} {Generalized {H}ouseholder
  transformations},\ }\href {https://doi.org/10.3390/e23050519} {\bibfield
  {journal} {\bibinfo  {journal} {Entropy}\ }\textbf {\bibinfo {volume} {23}},\
  \bibinfo {pages} {429} (\bibinfo {year} {2022})},\ \Eprint
  {https://arxiv.org/abs/arXiv:2112.15206} {arXiv:2112.15206} \BibitemShut
  {NoStop}%
\bibitem [{\citenamefont {Klyachko}\ \emph {et~al.}(2008)\citenamefont
  {Klyachko}, \citenamefont {Can}, \citenamefont
  {Binicio\ifmmode~\breve{g}\else \u{g}\fi{}lu},\ and\ \citenamefont
  {Shumovsky}}]{Klyachko-2008}%
  \BibitemOpen
  \bibfield  {author} {\bibinfo {author} {\bibfnamefont {A.~A.}\ \bibnamefont
  {Klyachko}}, \bibinfo {author} {\bibfnamefont {M.~A.}\ \bibnamefont {Can}},
  \bibinfo {author} {\bibfnamefont {S.}~\bibnamefont
  {Binicio\ifmmode~\breve{g}\else \u{g}\fi{}lu}},\ and\ \bibinfo {author}
  {\bibfnamefont {A.~S.}\ \bibnamefont {Shumovsky}},\ }\bibfield  {title}
  {\bibinfo {title} {Simple test for hidden variables in spin-1 systems},\
  }\href {https://doi.org/10.1103/PhysRevLett.101.020403} {\bibfield  {journal}
  {\bibinfo  {journal} {Physical Review Letters}\ }\textbf {\bibinfo {volume}
  {101}},\ \bibinfo {pages} {020403} (\bibinfo {year} {2008})},\ \Eprint
  {https://arxiv.org/abs/arXiv:0706.0126} {arXiv:0706.0126} \BibitemShut
  {NoStop}%
\bibitem [{\citenamefont {Cabello}(2008)}]{cabello:210401}%
  \BibitemOpen
  \bibfield  {author} {\bibinfo {author} {\bibfnamefont {A.}~\bibnamefont
  {Cabello}},\ }\bibfield  {title} {\bibinfo {title} {Experimentally testable
  state-independent quantum contextuality},\ }\href
  {https://doi.org/10.1103/PhysRevLett.101.210401} {\bibfield  {journal}
  {\bibinfo  {journal} {Physical Review Letters}\ }\textbf {\bibinfo {volume}
  {101}},\ \bibinfo {eid} {210401} (\bibinfo {year} {2008})},\ \Eprint
  {https://arxiv.org/abs/arXiv:0808.2456} {arXiv:0808.2456} \BibitemShut
  {NoStop}%
\bibitem [{\citenamefont {Pitowsky}(1994)}]{Pit-94}%
  \BibitemOpen
  \bibfield  {author} {\bibinfo {author} {\bibfnamefont {I.}~\bibnamefont
  {Pitowsky}},\ }\bibfield  {title} {\bibinfo {title} {{G}eorge {B}oole's
  `conditions of possible experience' and the quantum puzzle},\ }\href
  {https://doi.org/10.1093/bjps/45.1.95} {\bibfield  {journal} {\bibinfo
  {journal} {The British Journal for the Philosophy of Science}\ }\textbf
  {\bibinfo {volume} {45}},\ \bibinfo {pages} {95} (\bibinfo {year}
  {1994})}\BibitemShut {NoStop}%
\bibitem [{\citenamefont {Ziegler}(1994)}]{ziegler}%
  \BibitemOpen
  \bibfield  {author} {\bibinfo {author} {\bibfnamefont {G.~M.}\ \bibnamefont
  {Ziegler}},\ }\href {https://doi.org/10.1007/978-1-4613-8431-1} {\emph
  {\bibinfo {title} {Lectures on Polytopes}}},\ \bibinfo {series} {Graduate
  Texts in Mathematics}, Vol.\ \bibinfo {volume} {152}\ (\bibinfo  {publisher}
  {Springer},\ \bibinfo {address} {New York},\ \bibinfo {year}
  {1994})\BibitemShut {NoStop}%
\bibitem [{\citenamefont {Froissart}(1981)}]{froissart-81}%
  \BibitemOpen
  \bibfield  {author} {\bibinfo {author} {\bibfnamefont {M.}~\bibnamefont
  {Froissart}},\ }\bibfield  {title} {\bibinfo {title} {Constructive
  generalization of {B}ell's inequalities},\ }\href
  {https://doi.org/10.1007/BF02903286} {\bibfield  {journal} {\bibinfo
  {journal} {Il Nuovo Cimento B (11, 1971-1996)}\ }\textbf {\bibinfo {volume}
  {64}},\ \bibinfo {pages} {241} (\bibinfo {year} {1981})}\BibitemShut
  {NoStop}%
\bibitem [{\citenamefont {Pitowsky}(1986)}]{pitowsky-86}%
  \BibitemOpen
  \bibfield  {author} {\bibinfo {author} {\bibfnamefont {I.}~\bibnamefont
  {Pitowsky}},\ }\bibfield  {title} {\bibinfo {title} {The range of quantum
  probability},\ }\href {https://doi.org/10.1063/1.527066} {\bibfield
  {journal} {\bibinfo  {journal} {Journal of Mathematical Physics}\ }\textbf
  {\bibinfo {volume} {27}},\ \bibinfo {pages} {1556} (\bibinfo {year}
  {1986})}\BibitemShut {NoStop}%
\bibitem [{\citenamefont {Svozil}(2001)}]{svozil-2001-cesena}%
  \BibitemOpen
  \bibfield  {author} {\bibinfo {author} {\bibfnamefont {K.}~\bibnamefont
  {Svozil}},\ }\href {https://doi.org/10.48550/arXiv.quant-ph/0012066}
  {\bibinfo {title} {On generalized probabilities: correlation polytopes for
  automaton logic and generalized urn models, extensions of quantum mechanics
  and parameter cheats}} (\bibinfo {year} {2001}),\ \Eprint
  {https://arxiv.org/abs/arXiv:quant-ph/0012066} {arXiv:quant-ph/0012066}
  \BibitemShut {NoStop}%
\bibitem [{\citenamefont {Fukuda}(2024)}]{cdd-pck}%
  \BibitemOpen
  \bibfield  {author} {\bibinfo {author} {\bibfnamefont {K.}~\bibnamefont
  {Fukuda}},\ }\href {http://www.inf.ethz.ch/personal/fukudak/cdd\_home/}
  {\bibinfo {title} {{cdd}, {cddplus} and {cddlibib} packages}} (\bibinfo
  {year} {2024}),\ \bibinfo {note} {last update on October 21, 2022, accessed
  on July 11th, 2024}\BibitemShut {NoStop}%
\bibitem [{\citenamefont {Troffaes}(2020)}]{pycddlib}%
  \BibitemOpen
  \bibfield  {author} {\bibinfo {author} {\bibfnamefont {M.}~\bibnamefont
  {Troffaes}},\ }\href {https://pypi.org/project/pycddlib/} {\bibinfo {title}
  {{pycddlib} package}} (\bibinfo {year} {2020}),\ \bibinfo {note} {{P}ython
  wrapper for {K}omei {F}ukuda's {cddlib}, accessed on December 22th,
  2020}\BibitemShut {NoStop}%
\bibitem [{\citenamefont {von Neumann}(1931)}]{v-neumann-31}%
  \BibitemOpen
  \bibfield  {author} {\bibinfo {author} {\bibfnamefont {J.}~\bibnamefont {von
  Neumann}},\ }\bibfield  {title} {\bibinfo {title} {{\"{U}}ber {F}unktionen
  von {F}unktionaloperatoren},\ }\href {https://doi.org/10.2307/1968185}
  {\bibfield  {journal} {\bibinfo  {journal} {Annalen der Mathematik (Annals of
  Mathematics)}\ }\textbf {\bibinfo {volume} {32}},\ \bibinfo {pages} {191}
  (\bibinfo {year} {1931})}\BibitemShut {NoStop}%
\bibitem [{\citenamefont {Halmos}(1958)}]{halmos-vs}%
  \BibitemOpen
  \bibfield  {author} {\bibinfo {author} {\bibfnamefont {P.~R.}\ \bibnamefont
  {Halmos}},\ }\href {https://doi.org/10.1007/978-1-4612-6387-6} {\emph
  {\bibinfo {title} {Finite-Dimensional Vector Spaces}}},\ Undergraduate Texts
  in Mathematics\ (\bibinfo  {publisher} {Springer},\ \bibinfo {address} {New
  York},\ \bibinfo {year} {1958})\BibitemShut {NoStop}%
\bibitem [{\citenamefont {Shekarriz}\ and\ \citenamefont
  {Svozil}(2022)}]{svozil-2021-chroma}%
  \BibitemOpen
  \bibfield  {author} {\bibinfo {author} {\bibfnamefont {M.~H.}\ \bibnamefont
  {Shekarriz}}\ and\ \bibinfo {author} {\bibfnamefont {K.}~\bibnamefont
  {Svozil}},\ }\bibfield  {title} {\bibinfo {title} {Noncontextual coloring of
  orthogonality hypergraphs},\ }\href {https://doi.org/10.1063/5.0062801}
  {\bibfield  {journal} {\bibinfo  {journal} {Journal of Mathematical Physics}\
  }\textbf {\bibinfo {volume} {63}},\ \bibinfo {pages} {032104} (\bibinfo
  {year} {2022})},\ \Eprint {https://arxiv.org/abs/arXiv:2105.08520}
  {arXiv:2105.08520} \BibitemShut {NoStop}%
\bibitem [{\citenamefont {Svozil}(2025)}]{svozil-2025-color}%
  \BibitemOpen
  \bibfield  {author} {\bibinfo {author} {\bibfnamefont {K.}~\bibnamefont
  {Svozil}},\ }\bibfield  {title} {\bibinfo {title} {Chromatic quantum
  contextuality},\ }\href {https://doi.org/10.3390/e27040387} {\bibfield
  {journal} {\bibinfo  {journal} {Entropy}\ }\textbf {\bibinfo {volume} {27}},\
  \bibinfo {pages} {387} (\bibinfo {year} {2025})},\ \Eprint
  {https://arxiv.org/abs/arXiv:2501.15261} {arXiv:2501.15261} \BibitemShut
  {NoStop}%
\bibitem [{\citenamefont {Zierler}\ and\ \citenamefont
  {Schlessinger}(1975)}]{Zierler1975}%
  \BibitemOpen
  \bibfield  {author} {\bibinfo {author} {\bibfnamefont {N.}~\bibnamefont
  {Zierler}}\ and\ \bibinfo {author} {\bibfnamefont {M.}~\bibnamefont
  {Schlessinger}},\ }\bibfield  {title} {\bibinfo {title} {Boolean embeddings
  of orthomodular sets and quantum logic},\ }in\ \href
  {https://doi.org/10.1007/978-94-010-1795-4_14} {\emph {\bibinfo {booktitle}
  {The Logico-Algebraic Approach to Quantum Mechanics: Volume {I}: Historical
  Evolution}}},\ \bibinfo {editor} {edited by\ \bibinfo {editor} {\bibfnamefont
  {C.~A.}\ \bibnamefont {Hooker}}}\ (\bibinfo  {publisher} {Springer
  Netherlands},\ \bibinfo {address} {Dordrecht, The Netherlands},\ \bibinfo
  {year} {1975})\ pp.\ \bibinfo {pages} {247--262}\BibitemShut {NoStop}%
\end{thebibliography}%
\ifws

\bibliographystyle{spmpsci}

\else

\fi

\end{document}